\documentclass[%
 reprint,
 amsmath,amssymb,
 aps,
]{revtex4-2}

\usepackage{graphicx}
\usepackage{dcolumn}
\usepackage{bm}
\usepackage{amsmath}
\usepackage{physics}
\usepackage{ragged2e}
\usepackage{lipsum}
\usepackage[T1]{fontenc}
\usepackage{hyperref}
\usepackage{cleveref}
\usepackage{soul}
\usepackage{dsfont}
\usepackage{diagbox}
\usepackage{amsthm}
\usepackage{amsfonts}
\usepackage{xcolor}
\usepackage[export]{adjustbox}
\usepackage{pdfpages}
\usepackage{pgffor}

\makeatletter
\AtBeginDocument{\let\LS@rot\@undefined}
\makeatother

\begin{document}

\preprint{APS/123-QED}

\author{Jérémie Boudreault$^1$}

\author{Ross Shillito$^2$}

\author{Jean-Baptiste Bertrand$^1$}

\author{Baptiste Royer$^1$}

\affiliation{$^1$Départment de Physique and Institut Quantique, Université de Sherbrooke, Sherbrooke J1K 2R1 Québec, CAN}

\affiliation{$^2$Nord Quantique, Sherbrooke J1K 2X8 Québec, CAN}

\title{Using a Kerr interaction for GKP magic state preparation}

\begin{abstract}
Magic state distillation and injection is a promising strategy towards universal fault tolerant quantum computation, especially in architectures based on the bosonic Gottesman-Kitaev-Preskill (GKP) codes where non-Clifford gates remain challenging to implement. Here we address GKP magic state preparation by studying a non-Gaussian unitary mediated by a Kerr interaction which realizes a logical gate $\sqrt{H}_L$ for square GKP codes. This gate does not directly involve an auxiliary qubit and is compatible with finite energy constraints on the code. Fidelity can be further enhanced using the small-Big-small (SBS) error correction protocol and post-selection, making the scheme robust against a single photon loss event. We finally propose a circuit QED implementation to operate the Kerr interaction.
\end{abstract}

\maketitle

    \textit{Introduction---}A necessary step toward fault-tolerant universal computation is quantum error correction (QEC), which consists of redundantly encoding logical information in a physical system in order to protect it against errors. This error protection can be achieved in the infinite Hilbert space of bosonic systems such as motional modes of trapped ions \cite{Fluhmann_2019,Saffman_2010,Neeve_2022}, microwave cavities \cite{Sivak_2023, Quirion-2024, Grismo_2021, Lescanne_2020, Grimm_2020} or optical photonic systems \cite{Larsen_2021,Bourassa_2021,Tzitrin_2021,Hastrup_2022}. We focus on Gottesman-Kitaev-Preskill (GKP) codes \cite{Gottesman_2001} which are of interest partly due to their favorable performance when subjected to the damping error channel compared to other bosonic codes \cite{Albert}. Altough the approach we propose could be realized in different platforms, we more specifically consider strategies adapted to microwave-cavity systems, where photon loss is the dominant error model~\cite{Michael}. Microwave GKP states, which have already been stabilized and controlled in the laboratory \cite{Campagne_Ibarcq_2020,Sivak_2023, Quirion-2024}, are a particularly promising approach due to their extended lifetimes \cite{Sivak_2023} and their strong coupling to superconducting qubits which enables universal control and readout \cite{Eickbusch-2021,Heeres-2016,Didier_2015}. However, directly using the auxiliary qubit for logical operations limits the attainable fidelity due to its finite lifetime and leakage to higher levels~\cite{Hu_2019,Campagne_Ibarcq_2020, Rosenblum_2018}. It is therefore advantageous to look for and prioritize qubit-independent logical operations. In principle, logical GKP Clifford gates can be realized in a robust manner using Gaussian operations, but non-Clifford gates require higher-order interactions that are more challenging to realize. 
    Magic state preparation and injection is a promising alternative, but preparation protocols suffer from the same qubit limitations \cite{Campagne_Ibarcq_2020}, require long sequences \cite{Singh_2025}, or are intrinsically incompatible with the finite energy constraints of the GKP states \cite{Hastrup_2021}. Homodyne measurements with Gaussian operations are sufficient to prepare distillable magic states if individual GKP code words can be prepared \cite{Baragiola_2019}. However, to the best of our knowledge, a high-efficiency homodyne measurement of a microwave cavity mode has not yet been realized, motivating the search for alternative approaches.
\begin{figure}[h!]
    \includegraphics[width=1\linewidth, trim=2cm 0.5cm 0.5cm 0.5cm,clip]{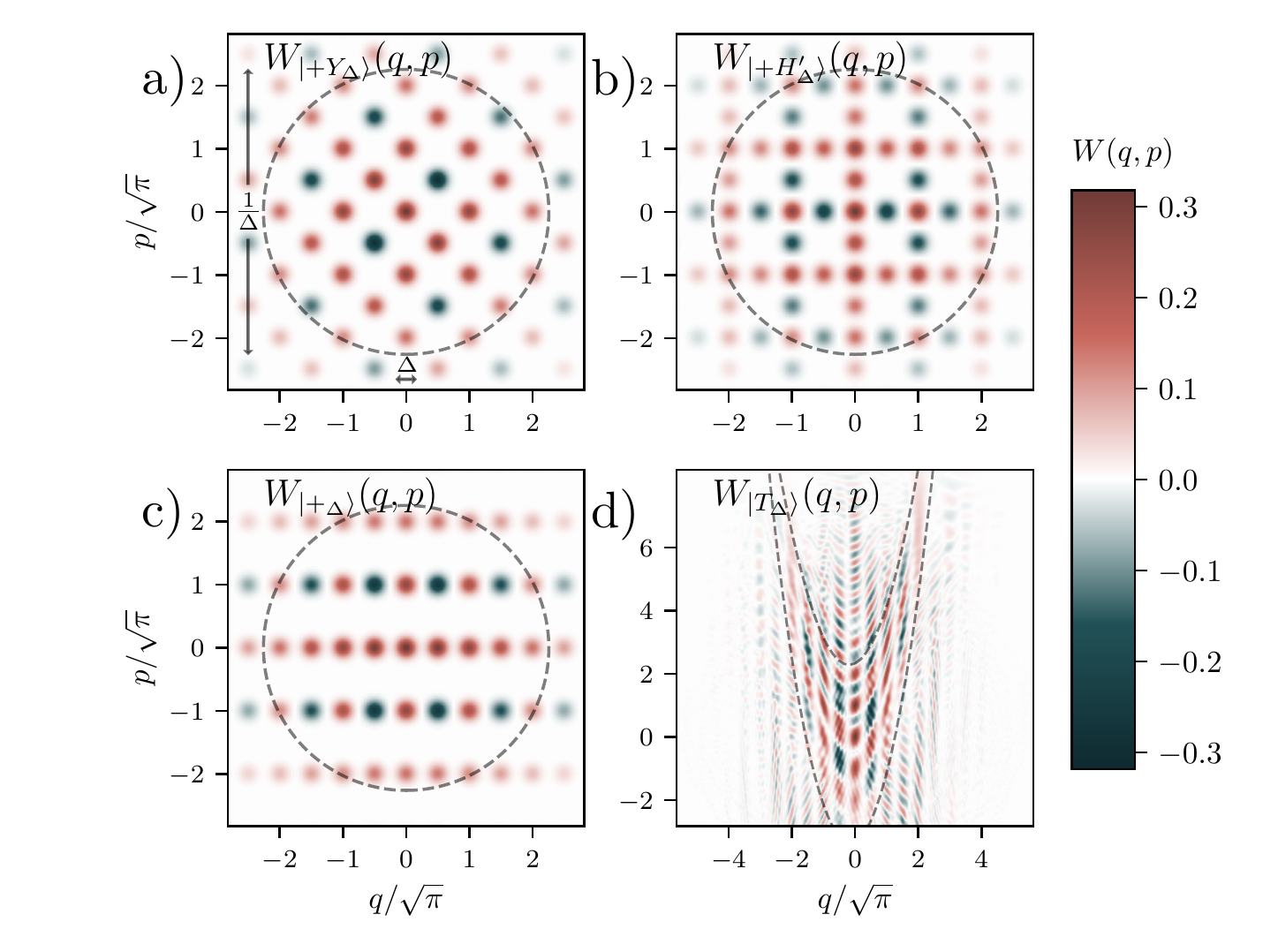}
    \caption{(a-b) Magic state preparation of $\ket{+H_\Delta'} = \sqrt{H}_L\ket{+Y_\Delta}$ using the Kerr unitary $\hat{U}_K$. (c-d) Magic state preparation of $\ket{T_\Delta} = T_L\ket{+_\Delta}$ using a cubic Hamiltonian. The dashed contour lines act as guides to the eye, indicating the shape of the envelope. The distorted shape in d) is computed using classical Hamilton's equations of motion for a cubic Hamiltonian (see SM). We set an envelope size of $\Delta=0.25$.
        }
    \label{fig:wigners}
\end{figure}

In this Letter, we study a logical non-Clifford gate $\sqrt{H}_L$ for square GKP states generated by a Kerr interaction. Since the evolution involves only the oscillator, the process is not directly limited by the control ancilla's lifetime. Moreover, it is suitable with finite energy states, as the gate commutes with the GKP envelope, and is therefore exact in the absence of errors in contrast to the cubic gate \cite{Gottesman_2001,Hastrup_2021}. Combined with error correction and post-selection, our approach provides a robust protocol for magic state preparation under photon loss. While this post-selection process requires an auxiliary qubit and therefore opens up the possibility of auxiliary qubit-induced logical errors, it has been shown that some QEC sequences are robust to phase ($T_2$) errors on the auxiliary qubit. As a result, the post-selection we propose could work well with a biased-noise qubit. \cite{Ding_2025}
In the following, we begin by summarizing the theory of ideal and finite GKP codes. We then describe the properties of the Kerr unitary and simulate magic state preparation under photon loss and with SBS error correction. Finally, we propose a circuit QED architecture that could operate the Kerr interaction.\\

\textit{GKP theory---}The ideal GKP states are +1 eigenstates of the stabilizers $\hat{S}_{q} = e^{i2\sqrt{\pi}\hat{q}}$ and $\hat{S}_{p} = e^{-i2\sqrt{\pi}\hat{p}}$ which are translation operators in the $\hat{p}$ and $\hat{q}$ quadrature coordinates, respectively. The ideal code words in the $\hat{q}$ basis are defined as Dirac combs $\ket{\mu_0}\propto \sum_{j\in\mathbb{Z}} \ket{2\sqrt{\pi}(j+\mu/2)}_q$ with $\mu \in \{0,1\}$, and the associated logical Pauli operators are given by ${X}_L = e^{-i\sqrt{\pi}\hat{p}}$ and ${Z}_L = e^{i\sqrt{\pi}\hat{q}}$.
The ideal states are not physical since they require infinite squeezing and energy. We thus consider the finite GKP states defined as
\begin{align}\label{eq:finiteEnergyStates}
    \ket{\mu_\Delta} = \mathcal{N}_\Delta \hat{E}_\Delta\ket{\mu_0},
\end{align}
where $\mathcal{N}_\Delta$ is a normalization constant and $\hat{E}_\Delta = \exp{-\Delta^2\hat{n}}$ is the envelope operator. To stabilize the code space, we consider the SBS protocol that accounts for the energy constraints during the error-correcting steps~\cite{Royer_2020,Neeve_2022,Campagne_Ibarcq_2020}. This protocol requires an auxiliary qubit and consists of a sequence of controlled displacements and qubit rotations. We describe the quantum channel induced by this protocol by the Kraus operators $\{\hat{K}_{jk}\}$, where $j,k \in \{g,e\}$ correspond to the result of the two measurements on the auxiliary qubit. A measurement result of $e$ indicates the presence of an error~\cite{Sivak_2023}. We can choose to express our states in the so-called (orthonormal) SBS basis $\{\ket{e,\mu}\}$~\cite{Hopfmueller_2024}, which enables us to make a subsystem decomposition of the physical bosonic Hilbert space, $\mathcal H_P$, as an error space and a logical qubit space, $\mathcal H_P = \mathcal H_E \otimes \mathcal H_L$. In this basis, a general bosonic state is then written as
\begin{align}
    \ket{\psi} = \sum_{e,\mu\in\{0,1\}}c_{e,\mu}\ket{e}\otimes\ket{\mu},
\end{align}
where $e = (e_q,e_p)$ is an error subspace label with two indices (see SM for more details). The no-error subspace, labeled by $(0,0)$, is a good approximation of the GKP code space defined from \cref{eq:finiteEnergyStates}, \emph{i.e.}, we have $\ket{\mu_\Delta} \approx \ket{(0, 0),\mu}$. In order to measure the fidelity of the logical information contained in a bosonic state, we map the latter to a qubit state $P:\mathcal{H}_P \to \mathcal{H}_L$ and compute the resulting logical fidelity. Below, we use the two following mappings resulting in different interpretations of fidelity:
\begin{align}\label{eq:perfectEDdecoding}
   \rho_{L_\Delta} &= \frac{\hat{P}_{\Delta}\rho\hat{P}_\Delta^\dag}{\Tr (\rho\hat{P}_\Delta)},
   \\
   \label{eq:SBSdecoding}
   \rho_{L_{\mathrm{sbs}}} = &\Tr_E(\rho) = \sum_{e}\hat{P}_e \rho \hat{P}^\dag_e,
\end{align}
where we have defined $\hat P_\Delta = \ket{0}\bra{0_\Delta} + \ket{1}\bra{1_\Delta}$~\footnote{Note that $\hat P_\Delta$ is not quite a projector since the code words defined by \cref{eq:finiteEnergyStates} are not perfectly orthogonal.} and $\hat{P}_{e} = \ket{0}\bra{e,0} + \ket{1}\bra{e,1}$. The first mapping above (\cref{eq:perfectEDdecoding}) provides an upper bound on fidelity by only retaining the logical information in the code space defined by \cref{eq:finiteEnergyStates}, which implies an unrealistic perfect error detecting protocol before decoding. We denote $\mathcal{F}_\Delta$ the fidelity obtained from comparing $\rho_{L_\Delta}$ with the target logical qubit state and define the success probability of the post-selection in this context as $\Tr(\rho\hat{P}_\Delta)$. More realistically, the actual decoding would follow a procedure closer to an experimental procedure. The second mapping above, given by \cref{eq:SBSdecoding}, averages the logical information contained in the error subspaces defined by the SBS basis and provides a more accurate description of the decoded qubit state. We denote $\mathcal{F}_{\mathrm{sbs}}$ the fidelity obtained from comparing $\rho_{\mathrm{sbs}}$ to the logical target qubit state. Below, we study how to increase $\mathcal{F}_{\mathrm{sbs}}$ by applying many rounds of SBS error correction and by post-selecting on measurement results where the auxiliary qubit is in state $g$. In this context, we define the success probability as the probability of applying $N$ rounds of SBS error correction without detecting any errors, \emph{i.e.} $\operatorname{Tr}[(\hat K_{gg}^\dag)^N\hat K_{gg}^N\rho]$.\\
\begin{figure*}[ht]
    \centering
    \includegraphics[width=0.98\linewidth]{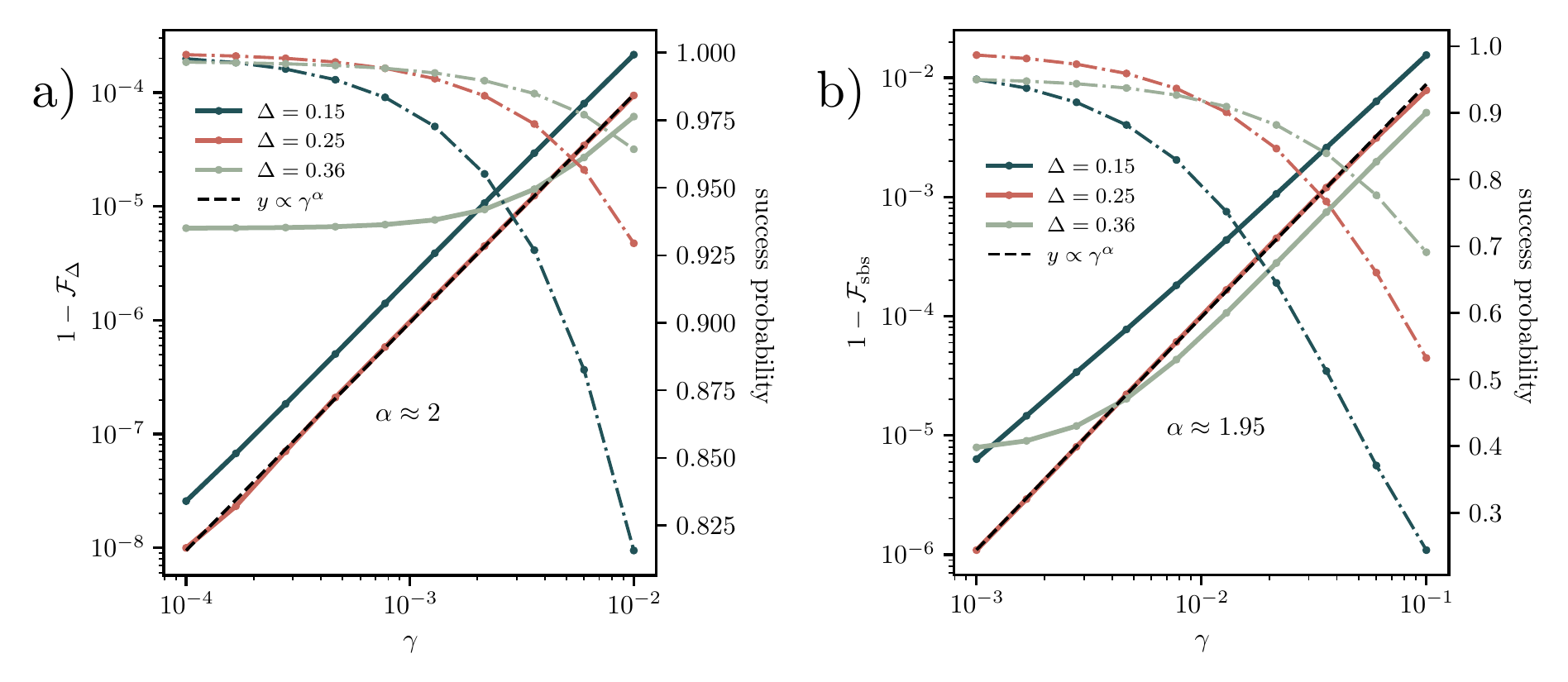}
    \caption{Magic state preparation under photon loss. a) Preparation of $\ket{+H'_\Delta} = \sqrt{H}_L\ket{+Y_\Delta}$ assuming perfect error detection. b) Preparation of $\ket{(0,0),H'}\approx \sqrt{H}_L\ket{(0,0),+Y}$ with $N=30$ rounds of SBS error correction and post-selection. Logical infidelity (solid lines) and the associated success probability (dashed lines) are plotted as a function of the loss parameter $\gamma = 1 - \exp(-\kappa t)$.}
    \label{fig:fidelities}
\end{figure*}

\textit{Magic state preparation with a Kerr unitary---}
While magic states are a promising solution to implement a non-Clifford gate for GKP codes, adopting the right strategy to prepare them is not obvious. Indeed, initial magic state's fidelity must surpass a certain threshold \cite{Bravyi_2005} and should be further optimized if one wants to minimize the already expensive overhead of distillation.
We now propose a new strategy to prepare a H-type \cite{Bravyi_2005} magic state
\begin{align}
    \ket{H'} = \sqrt{H}\ket{Y}
\end{align}
for GKP states. We denote the magic state $\ket{H'}$ as such since it's an image of the Hadamard eigenstates $\ket{\pm H}$ under Clifford gates. In the Bloch sphere, the gate $\sqrt{H}$ corresponds to a $\pi/2$ rotation around the axis defined by $(\theta=\pi/4, \phi=0)$. Its action on the Hadamard eigenstates is defined as $\sqrt{H}\ket{\pm H} = e^{i\pi(1\mp 1)/4}\ket{\pm H}$. 

For square GKP states, the Hadamard gate corresponds to the Fourier transform $H_L\equiv\hat{\mathcal{F}}=e^{i\pi\hat{n}/2}$ and its eigenstates have support only on specific Fock states $\ket{\pm H_\Delta} = \sum_{n\in\mathbb{N}} c_{\pm,n}\ket{4n + (1 \mp 1)}$. \cite{Gottesman_2001} This special support can be leveraged to implement a logical $\sqrt{H}_L$ gate. 
Explicitly, the desired unitary can be obtained by considering the square of the generator in the Hadamard gate, more precisely
\begin{align}
    \sqrt{H}_L \equiv \hat{U}_K = e^{i\pi\hat{n}^2/8},
\end{align}
which is generated by a quartic Kerr Hamiltonian, $\hat H = K\hat{n}^2/2$ (we use $\hbar = 1$ throughout this paper). The main feature of this approach is that the Gaussian envelope of the finite-energy GKP state is preserved throughout the process since $[\hat{n}^2,\hat{E}_\Delta] = 0$. In other words, this gate is exact in the absence of errors. In contrast, the logical gate $T_L = \ket{0}\bra{0} + e^{i\pi/4}\ket{1}\bra{1}$, generated by a cubic Hamiltonian for square GKP states~\cite{Gottesman_2001}, which can also be used to prepare magic states $\ket{T_\Delta} = T_L\ket{+_\Delta}$. This latter approach distorts the envelope since $[T_L,\hat{E}_\Delta] \neq 0$ and is therefore inherently inadequate~\cite{Hastrup_2021}.
Examples of errorless magic state preparation using these two methods are illustrated in figure \ref{fig:wigners}.
Another feature of $\hat{U}_K$ is that its evolution is qubit independent and consequently not directly affected by ancilla's relaxation errors.

\textit{Photon loss---}While the Kerr unitary $\hat{U}_K$ is exact in the code space, it is a non-Gaussian operator that amplifies errors as they occur. Indeed, during the evolution under the quartic Kerr Hamiltonian, the peaks of the Wigner function get distorted only to refocus themselves precisely at integer multiples of $t_{K} = \pi/4K$. We especially consider the case of photon loss which is the dominant source of errors for superconducting cavities. We introduce the dissipator $\mathcal{D}[\sqrt{\kappa}\hat{a}]$ with the superoperator $\mathcal{D}[\hat{L}](\rho) = \hat{L}\rho \hat{L}^\dag - \frac{1}{2}\{\hat{L}^\dag \hat{L}, \rho\}$. This error channel is characterized by a decay rate $\kappa$, and the loss parameter $\gamma = 1 - e^{-\kappa t_K}$ gives the loss probability after a time $t_K$ for each photon. The relation $\hat{U}_K\hat{a} = e^{i\pi/8}e^{-i\hat{n}\pi/4}\hat{a}\hat{U}_K$  highlights that a photon loss error before the gate transforms into a photon loss \emph{and} a large rotation which ruins the logical information. Importantly, however, a photon loss during the gate leaves the state outside of the code space, which means that such an error is still detectable even if it cannot be corrected. In the following, we therefore consider the gate $\hat U_K$ in the context of (offline) magic state preparation, where post-selection can be applied to eliminate cases where a single photon loss occurs.

\textit{Simulations---}We simulate the preparation of a GKP magic state $\ket{H'}$ with the Kerr unitary $\hat{U}_K$ under the photon loss channel in figure \ref{fig:fidelities}. In figure \ref{fig:fidelities}a), we focus on the code space fidelity of $\ket{H'_\Delta} = \sqrt{H}_L\ket{+Y_\Delta}$. The fidelity remains high, \emph{i.e.} $(\mathcal{F}_\Delta\gtrsim1 - 10^{-4})$ with a high success probability $(>80\%)$ for a wide range of GKP sizes. This shows that the logical information in the code space is generally well protected against loss. A major contributing factor is that the finite state envelope preserves its shape and its position centered at the origin throughout the process, which prevents an increase in photon loss and other spurious effects due to energy injection in the system. 
It also facilitates decoding as the physical fidelity with the states $\ket{\mu_\Delta}$ remains maximal. The high fidelity of prepared magic states can also be explained by the fact that states that underwent a single photon loss are detected and rejected. To incur a logical error, a less likely two-photon loss error is needed, which occurs with probability $\sim \gamma^2$. Photon loss itself is less likely when decreasing GKP size (increasing $\Delta$) since $\overline{n}\sim 1/(2\Delta^{2}) - 1/2$~\cite{Albert}, leading to worse fidelities when comparing $\Delta = 0.15$ (blue) with $\Delta = 0.25$ (red). However, the overlap between code words $\braket{0_\Delta}{1_\Delta}$ increases with increasing $\Delta$, which intrinsically limits the state preparation performance for $\Delta=0.36$ (olive green). The competition between photon loss probability and code words overlap explains why a GKP of larger size (smaller $\Delta$) performs better at very low loss, whereas a smaller GKP is preferable at higher loss. At really low loss rates $\gamma\lesssim10^{-4}$, though questionably attainable, we would expect $\Delta=0.15$ to outperform $\Delta=0.25$. Perfect error detection is also not achievable in practice, and figure \ref{fig:fidelities}a) should be interpreted as the fundamental limit of the Kerr unitary.

 To have a better picture, we consider in figure \ref{fig:fidelities}b) magic state preparation with more realistic approximate GKP code words $\ket{(0,0),\mu}$ in the SBS basis. The fidelity is also averaged over all error subspaces after $N=30$ rounds of SBS error correction and post-selection. The effect of $\hat{U}_K$ on the approximate GKP code words is similar to the desired logical gate, \emph{i.e.} $\hat{U}_K\ket{(0,0),\mu} \approx \sqrt{H}_L\ket{(0,0),\mu}$, but is different on the other error subspaces spanned by $\ket{e\neq(0,0),\mu}$ (see SM) because of their arbitrary support in the Fock basis. As a consequence, without post selection $\mathcal{F}_{\mathrm{sbs}}(\gamma) \leq \mathcal{F}_\Delta(\gamma)$. However, SBS error correction and post-selection rounds $\hat{K}_{gg}^N$ allow to retrieve the perfect error detection limit set by $\mathcal{F}_\Delta(\gamma)$. Indeed, the post-selection largely increases the probability of detecting a photon loss event, effectively increasing the no error subspace contribution to $\mathcal{F_{\mathrm{sbs}}}$. We reach the SBS protocol steady state when no further improvements are made, and it turns out that $N=30$ is sufficient (see SM). We can observe that the quadratic gain is mostly retrieved and that both fidelity and success probability are only lightly altered (note that the $x$-axis values of $\gamma$ differ between curves).  We attribute the early tendency to saturate of $\Delta=0.15$ to numerical errors and not to a physical effect as again higher GKP size should outperform those with smaller size at low loss.\\
Even though $30$ rounds of SBS error correction and post-selection might seem high and impractical to realize, considerable gains in fidelity can still be made with only a few rounds of post-selection (see SM), and we note that \ref{fig:fidelities}b) stands as an upper bound on what could be achieved in principle with the SBS error correcting procedure. For example, with a loss parameter of $\gamma = 10^{-3}$ a fidelity of $\mathcal{F}_{\mathrm{sbs}} \approx 2.4\times10^{-4}$ could be obtained with 2 rounds of post-selection. Both plots in figure \ref{fig:fidelities} together demonstrate the performance and the compatibility of the unitary $\hat{U}_K$ and the SBS protocol with realistic finite GKP states.

The analysis above did not include the effect of auxiliary measurement errors on the post-selection. We remark, however, that when using post-selection only false positive results (with probability $p(g|e)$) impact the fidelity. This can be mitigated at a small cost in success probability by, for example, exploring different thresholding strategies in dispersive readout \cite{Alexandre_RMP} that decrease $p(g|e)$ and increase $p(e|g)$.
Another simplification we made regarding state preparation in figure \ref{fig:fidelities} is that the initial state is noiseless. Considering all main error sources together, that is photon loss during initialization, gate evolution and error correction, including measurement errors during post-selection and using realistic values of $\Delta=0.36, \gamma = 10^{-2}, \mathrm{P}(e|g) = 10^{-3}$, we obtain a magic state fidelity $\mathcal{F}_{\mathrm{sbs}}\approx 0.996$ with a success probability of $\mathcal{P}\approx 81\%$. In the supplemental material, we provide an analysis of the magic state preparation fidelity that includes measurement errors, noisy initial states as well as oscillator dephasing.

\begin{figure}
    \centering
    \vspace{0.3cm}
    \includegraphics[width=1\linewidth]{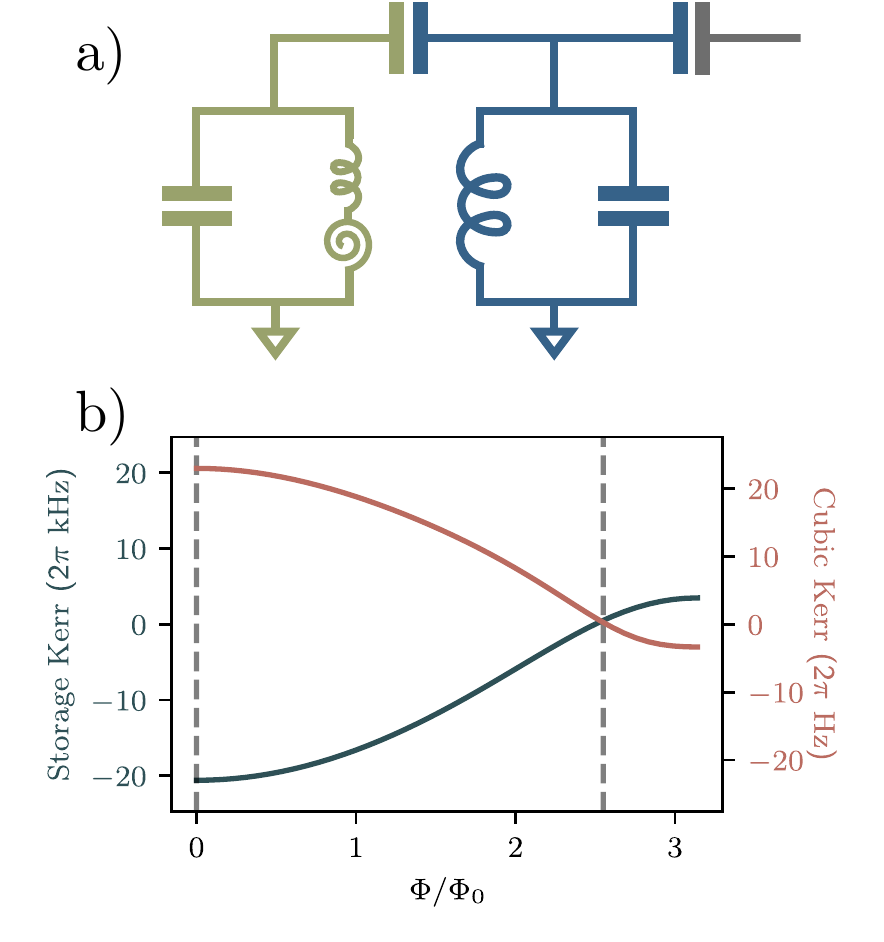}
    \caption{a) Circuit diagram for the proposed implementation, where the SNAIL circuit (green) is capacitively coupled to the storage mode (blue). b) Sweep of the induced nonlinearities in the storage mode as a function of the external flux bias, $\Phi_e$ (units of $\Phi_0 = h/2e$). For the proposed parameters we find two operating points -- for stabilization at $\Phi/\Phi_0 \approx 2.56$, and for the self-Kerr gate at $\Phi/\Phi_0 = 0$. Errors induced by the small cubic-Kerr term are largely detectable by SBS, and thus can be postselected.}
    \label{fig:circuit_and_Kerr}
\end{figure}

\textit{Circuit QED Implementation---}
The implementation of the target unitary $\hat U_K$ requires a tunable Kerr interaction. The ability to operate this feature was explored in many architectures such as SQUIDs \cite{Hua_2025, Zhang_2017}, quartons \cite{Ye_2021, Ye_2024}, hybrid cavity-qubit systems \cite{Elliott_2018} or hybrid SAW-(SQUID-array) systems \cite{Scigliuzzo_2025}. Here, we briefly study a possible implementation using a microwave cavity coupled dispersively to a flux-biased Superconducting Nonlinear Asymmetric Inductive eLement (SNAIL) \cite{Miano_2022, Sivak_2019, He_2023, Ranadive_2022}.
%via fast flux pulsing the SNAIL from its operating point.
An example lumped-element circuit is given in \Cref{fig:circuit_and_Kerr}a).  A similar technique was recently used to prepare Cat qubits via implementing a self-Kerr gate \cite{He2023}, where nonlinearities as high as $5$ MHz were achieved.
In choosing the appropriate SNAIL parameters, it is important to consider higher-order induced nonlinearities, since these can be particularly impactful for higher-energy GKP states.  We additionally require a dispersive coupling to the SNAIL to ensure the storage retains a high Q-factor and is not Purcell limited. We find that with an appropriate choice of parameters, we can induce a self-Kerr of  $K/2\pi \approx -20$kHz, leading to a corresponding gate time of $6.3 \mu\text{s}$.  In this regime, the higher-order nonlinearities are limited, and have little impact on the fidelity of the prepared magic states (see SM). For a cavity lifetime of $610\mu\text{s}$, that would correspond to $\gamma \approx 1.07\times 10^{-2}$. 

\textit{Conclusion---}We have introduced a qubit-independent protocol for GKP magic state preparation with a Kerr unitary which preserves the envelope of finite energy states. This procedure can be made robust to single photon loss errors during offline preparation through SBS error correction and post-selection using a biased noise ancilla. While we focused on microwave-based platforms, the preparation protocol we suggested is compatible with different experimental platforms such as microwave cavities, motional modes of trapped ions, or acoustic resonators. We gave a possible realization with a cavity-SNAIL system where higher-order Kerr nonlinearities are very small, but many oscillator-qubit architectures supporting a tunable Kerr interaction remain to be explored. This work provides a pathway towards what could be a robust and potentially low-cost solution to practical GKP magic state preparation.

\section*{Acknowledgements}
This work was funded by the Army Research Office under the grant W911NF2310045, NSERC, the Fonds de recherche du Québec – Nature et technologie and the Canada First Research Excellence Fund.

\bibliography{apssamp}% Produces the bibliography via BibTeX.

%apsrev4-2.bst 2019-01-14 (MD) hand-edited version of apsrev4-1.bst
%Control: key (0)
%Control: author (8) initials jnrlst
%Control: editor formatted (1) identically to author
%Control: production of article title (0) allowed
%Control: page (0) single
%Control: year (1) truncated
%Control: production of eprint (0) enabled
\begin{thebibliography}{41}%
\makeatletter
\providecommand \@ifxundefined [1]{%
 \@ifx{#1\undefined}
}%
\providecommand \@ifnum [1]{%
 \ifnum #1\expandafter \@firstoftwo
 \else \expandafter \@secondoftwo
 \fi
}%
\providecommand \@ifx [1]{%
 \ifx #1\expandafter \@firstoftwo
 \else \expandafter \@secondoftwo
 \fi
}%
\providecommand \natexlab [1]{#1}%
\providecommand \enquote  [1]{``#1''}%
\providecommand \bibnamefont  [1]{#1}%
\providecommand \bibfnamefont [1]{#1}%
\providecommand \citenamefont [1]{#1}%
\providecommand \href@noop [0]{\@secondoftwo}%
\providecommand \href [0]{\begingroup \@sanitize@url \@href}%
\providecommand \@href[1]{\@@startlink{#1}\@@href}%
\providecommand \@@href[1]{\endgroup#1\@@endlink}%
\providecommand \@sanitize@url [0]{\catcode `\\12\catcode `\$12\catcode `\&12\catcode `\#12\catcode `\^12\catcode `\_12\catcode `\%12\relax}%
\providecommand \@@startlink[1]{}%
\providecommand \@@endlink[0]{}%
\providecommand \url  [0]{\begingroup\@sanitize@url \@url }%
\providecommand \@url [1]{\endgroup\@href {#1}{\urlprefix }}%
\providecommand \urlprefix  [0]{URL }%
\providecommand \Eprint [0]{\href }%
\providecommand \doibase [0]{https://doi.org/}%
\providecommand \selectlanguage [0]{\@gobble}%
\providecommand \bibinfo  [0]{\@secondoftwo}%
\providecommand \bibfield  [0]{\@secondoftwo}%
\providecommand \translation [1]{[#1]}%
\providecommand \BibitemOpen [0]{}%
\providecommand \bibitemStop [0]{}%
\providecommand \bibitemNoStop [0]{.\EOS\space}%
\providecommand \EOS [0]{\spacefactor3000\relax}%
\providecommand \BibitemShut  [1]{\csname bibitem#1\endcsname}%
\let\auto@bib@innerbib\@empty
%</preamble>
\bibitem [{\citenamefont {Flühmann}\ \emph {et~al.}(2019)\citenamefont {Flühmann}, \citenamefont {Nguyen}, \citenamefont {Marinelli}, \citenamefont {Negnevitsky}, \citenamefont {Mehta},\ and\ \citenamefont {Home}}]{Fluhmann_2019}%
  \BibitemOpen
  \bibfield  {author} {\bibinfo {author} {\bibfnamefont {C.}~\bibnamefont {Flühmann}}, \bibinfo {author} {\bibfnamefont {T.~L.}\ \bibnamefont {Nguyen}}, \bibinfo {author} {\bibfnamefont {M.}~\bibnamefont {Marinelli}}, \bibinfo {author} {\bibfnamefont {V.}~\bibnamefont {Negnevitsky}}, \bibinfo {author} {\bibfnamefont {K.}~\bibnamefont {Mehta}},\ and\ \bibinfo {author} {\bibfnamefont {J.~P.}\ \bibnamefont {Home}},\ }\bibfield  {title} {\bibinfo {title} {Encoding a qubit in a trapped-ion mechanical oscillator},\ }\href {https://doi.org/10.1038/s41586-019-0960-6} {\bibfield  {journal} {\bibinfo  {journal} {Nature}\ }\textbf {\bibinfo {volume} {566}},\ \bibinfo {pages} {513–517} (\bibinfo {year} {2019})}\BibitemShut {NoStop}%
\bibitem [{\citenamefont {Saffman}\ \emph {et~al.}(2010)\citenamefont {Saffman}, \citenamefont {Walker},\ and\ \citenamefont {M\o{}lmer}}]{Saffman_2010}%
  \BibitemOpen
  \bibfield  {author} {\bibinfo {author} {\bibfnamefont {M.}~\bibnamefont {Saffman}}, \bibinfo {author} {\bibfnamefont {T.~G.}\ \bibnamefont {Walker}},\ and\ \bibinfo {author} {\bibfnamefont {K.}~\bibnamefont {M\o{}lmer}},\ }\bibfield  {title} {\bibinfo {title} {Quantum information with rydberg atoms},\ }\href {https://doi.org/10.1103/RevModPhys.82.2313} {\bibfield  {journal} {\bibinfo  {journal} {Rev. Mod. Phys.}\ }\textbf {\bibinfo {volume} {82}},\ \bibinfo {pages} {2313} (\bibinfo {year} {2010})}\BibitemShut {NoStop}%
\bibitem [{\citenamefont {{de Neeve}}\ \emph {et~al.}(2022)\citenamefont {{de Neeve}}, \citenamefont {{Nguyen}}, \citenamefont {{Behrle}},\ and\ \citenamefont {{Home}}}]{Neeve_2022}%
  \BibitemOpen
  \bibfield  {author} {\bibinfo {author} {\bibfnamefont {B.}~\bibnamefont {{de Neeve}}}, \bibinfo {author} {\bibfnamefont {T.-L.}\ \bibnamefont {{Nguyen}}}, \bibinfo {author} {\bibfnamefont {T.}~\bibnamefont {{Behrle}}},\ and\ \bibinfo {author} {\bibfnamefont {J.~P.}\ \bibnamefont {{Home}}},\ }\bibfield  {title} {\bibinfo {title} {{Error correction of a logical grid state qubit by dissipative pumping}},\ }\href {https://doi.org/10.1038/s41567-021-01487-7} {\bibfield  {journal} {\bibinfo  {journal} {Nature Physics}\ }\textbf {\bibinfo {volume} {18}},\ \bibinfo {pages} {296} (\bibinfo {year} {2022})}\BibitemShut {NoStop}%
\bibitem [{\citenamefont {Sivak}\ \emph {et~al.}(2023)\citenamefont {Sivak}, \citenamefont {Eickbusch}, \citenamefont {Royer}, \citenamefont {Singh}, \citenamefont {Tsioutsios}, \citenamefont {Ganjam}, \citenamefont {Miano}, \citenamefont {Brock}, \citenamefont {Ding}, \citenamefont {Frunzio}, \citenamefont {Girvin}, \citenamefont {Schoelkopf},\ and\ \citenamefont {Devoret}}]{Sivak_2023}%
  \BibitemOpen
  \bibfield  {author} {\bibinfo {author} {\bibfnamefont {V.~V.}\ \bibnamefont {Sivak}}, \bibinfo {author} {\bibfnamefont {A.}~\bibnamefont {Eickbusch}}, \bibinfo {author} {\bibfnamefont {B.}~\bibnamefont {Royer}}, \bibinfo {author} {\bibfnamefont {S.}~\bibnamefont {Singh}}, \bibinfo {author} {\bibfnamefont {I.}~\bibnamefont {Tsioutsios}}, \bibinfo {author} {\bibfnamefont {S.}~\bibnamefont {Ganjam}}, \bibinfo {author} {\bibfnamefont {A.}~\bibnamefont {Miano}}, \bibinfo {author} {\bibfnamefont {B.~L.}\ \bibnamefont {Brock}}, \bibinfo {author} {\bibfnamefont {A.~Z.}\ \bibnamefont {Ding}}, \bibinfo {author} {\bibfnamefont {L.}~\bibnamefont {Frunzio}}, \bibinfo {author} {\bibfnamefont {S.~M.}\ \bibnamefont {Girvin}}, \bibinfo {author} {\bibfnamefont {R.~J.}\ \bibnamefont {Schoelkopf}},\ and\ \bibinfo {author} {\bibfnamefont {M.~H.}\ \bibnamefont {Devoret}},\ }\bibfield  {title} {\bibinfo {title} {Real-time quantum error correction beyond break-even},\ }\href {https://doi.org/10.1038/s41586-023-05782-6} {\bibfield
  {journal} {\bibinfo  {journal} {Nature}\ }\textbf {\bibinfo {volume} {616}},\ \bibinfo {pages} {50–55} (\bibinfo {year} {2023})}\BibitemShut {NoStop}%
\bibitem [{\citenamefont {Lachance-Quirion}\ \emph {et~al.}(2024)\citenamefont {Lachance-Quirion}, \citenamefont {Lemonde}, \citenamefont {Simoneau}, \citenamefont {St-Jean}, \citenamefont {Lemieux}, \citenamefont {Turcotte}, \citenamefont {Wright}, \citenamefont {Lacroix}, \citenamefont {Fr\'echette-Viens}, \citenamefont {Shillito}, \citenamefont {Hopfmueller}, \citenamefont {Tremblay}, \citenamefont {Frattini}, \citenamefont {Camirand~Lemyre},\ and\ \citenamefont {St-Jean}}]{Quirion-2024}%
  \BibitemOpen
  \bibfield  {author} {\bibinfo {author} {\bibfnamefont {D.}~\bibnamefont {Lachance-Quirion}}, \bibinfo {author} {\bibfnamefont {M.-A.}\ \bibnamefont {Lemonde}}, \bibinfo {author} {\bibfnamefont {J.~O.}\ \bibnamefont {Simoneau}}, \bibinfo {author} {\bibfnamefont {L.}~\bibnamefont {St-Jean}}, \bibinfo {author} {\bibfnamefont {P.}~\bibnamefont {Lemieux}}, \bibinfo {author} {\bibfnamefont {S.}~\bibnamefont {Turcotte}}, \bibinfo {author} {\bibfnamefont {W.}~\bibnamefont {Wright}}, \bibinfo {author} {\bibfnamefont {A.}~\bibnamefont {Lacroix}}, \bibinfo {author} {\bibfnamefont {J.}~\bibnamefont {Fr\'echette-Viens}}, \bibinfo {author} {\bibfnamefont {R.}~\bibnamefont {Shillito}}, \bibinfo {author} {\bibfnamefont {F.}~\bibnamefont {Hopfmueller}}, \bibinfo {author} {\bibfnamefont {M.}~\bibnamefont {Tremblay}}, \bibinfo {author} {\bibfnamefont {N.~E.}\ \bibnamefont {Frattini}}, \bibinfo {author} {\bibfnamefont {J.}~\bibnamefont {Camirand~Lemyre}},\ and\ \bibinfo {author} {\bibfnamefont {P.}~\bibnamefont {St-Jean}},\
  }\bibfield  {title} {\bibinfo {title} {Autonomous quantum error correction of gottesman-kitaev-preskill states},\ }\href {https://doi.org/10.1103/PhysRevLett.132.150607} {\bibfield  {journal} {\bibinfo  {journal} {Phys. Rev. Lett.}\ }\textbf {\bibinfo {volume} {132}},\ \bibinfo {pages} {150607} (\bibinfo {year} {2024})}\BibitemShut {NoStop}%
\bibitem [{\citenamefont {Grimsmo}\ and\ \citenamefont {Puri}(2021)}]{Grismo_2021}%
  \BibitemOpen
  \bibfield  {author} {\bibinfo {author} {\bibfnamefont {A.~L.}\ \bibnamefont {Grimsmo}}\ and\ \bibinfo {author} {\bibfnamefont {S.}~\bibnamefont {Puri}},\ }\bibfield  {title} {\bibinfo {title} {Quantum error correction with the gottesman-kitaev-preskill code},\ }\href {https://doi.org/10.1103/PRXQuantum.2.020101} {\bibfield  {journal} {\bibinfo  {journal} {PRX Quantum}\ }\textbf {\bibinfo {volume} {2}},\ \bibinfo {pages} {020101} (\bibinfo {year} {2021})}\BibitemShut {NoStop}%
\bibitem [{\citenamefont {Lescanne}\ \emph {et~al.}(2020)\citenamefont {Lescanne}, \citenamefont {Villiers}, \citenamefont {Peronnin}, \citenamefont {Sarlette}, \citenamefont {Delbecq}, \citenamefont {Huard}, \citenamefont {Kontos}, \citenamefont {Mirrahimi},\ and\ \citenamefont {Leghtas}}]{Lescanne_2020}%
  \BibitemOpen
  \bibfield  {author} {\bibinfo {author} {\bibfnamefont {R.}~\bibnamefont {Lescanne}}, \bibinfo {author} {\bibfnamefont {M.}~\bibnamefont {Villiers}}, \bibinfo {author} {\bibfnamefont {T.}~\bibnamefont {Peronnin}}, \bibinfo {author} {\bibfnamefont {A.}~\bibnamefont {Sarlette}}, \bibinfo {author} {\bibfnamefont {M.}~\bibnamefont {Delbecq}}, \bibinfo {author} {\bibfnamefont {B.}~\bibnamefont {Huard}}, \bibinfo {author} {\bibfnamefont {T.}~\bibnamefont {Kontos}}, \bibinfo {author} {\bibfnamefont {M.}~\bibnamefont {Mirrahimi}},\ and\ \bibinfo {author} {\bibfnamefont {Z.}~\bibnamefont {Leghtas}},\ }\bibfield  {title} {\bibinfo {title} {Exponential suppression of bit-flips in a qubit encoded in an oscillator},\ }\href {https://doi.org/10.1038/s41567-020-0824-x} {\bibfield  {journal} {\bibinfo  {journal} {Nature Physics}\ }\textbf {\bibinfo {volume} {16}},\ \bibinfo {pages} {509–513} (\bibinfo {year} {2020})}\BibitemShut {NoStop}%
\bibitem [{\citenamefont {Grimm}\ \emph {et~al.}(2020)\citenamefont {Grimm}, \citenamefont {Frattini}, \citenamefont {Puri}, \citenamefont {Mundhada}, \citenamefont {Touzard}, \citenamefont {Mirrahimi}, \citenamefont {Girvin}, \citenamefont {Shankar},\ and\ \citenamefont {Devoret}}]{Grimm_2020}%
  \BibitemOpen
  \bibfield  {author} {\bibinfo {author} {\bibfnamefont {A.}~\bibnamefont {Grimm}}, \bibinfo {author} {\bibfnamefont {N.~E.}\ \bibnamefont {Frattini}}, \bibinfo {author} {\bibfnamefont {S.}~\bibnamefont {Puri}}, \bibinfo {author} {\bibfnamefont {S.~O.}\ \bibnamefont {Mundhada}}, \bibinfo {author} {\bibfnamefont {S.}~\bibnamefont {Touzard}}, \bibinfo {author} {\bibfnamefont {M.}~\bibnamefont {Mirrahimi}}, \bibinfo {author} {\bibfnamefont {S.~M.}\ \bibnamefont {Girvin}}, \bibinfo {author} {\bibfnamefont {S.}~\bibnamefont {Shankar}},\ and\ \bibinfo {author} {\bibfnamefont {M.~H.}\ \bibnamefont {Devoret}},\ }\bibfield  {title} {\bibinfo {title} {Stabilization and operation of a kerr-cat qubit},\ }\href {https://doi.org/10.1038/s41586-020-2587-z} {\bibfield  {journal} {\bibinfo  {journal} {Nature}\ }\textbf {\bibinfo {volume} {584}},\ \bibinfo {pages} {205–209} (\bibinfo {year} {2020})}\BibitemShut {NoStop}%
\bibitem [{\citenamefont {Larsen}\ \emph {et~al.}(2021)\citenamefont {Larsen}, \citenamefont {Chamberland}, \citenamefont {Noh}, \citenamefont {Neergaard-Nielsen},\ and\ \citenamefont {Andersen}}]{Larsen_2021}%
  \BibitemOpen
  \bibfield  {author} {\bibinfo {author} {\bibfnamefont {M.~V.}\ \bibnamefont {Larsen}}, \bibinfo {author} {\bibfnamefont {C.}~\bibnamefont {Chamberland}}, \bibinfo {author} {\bibfnamefont {K.}~\bibnamefont {Noh}}, \bibinfo {author} {\bibfnamefont {J.~S.}\ \bibnamefont {Neergaard-Nielsen}},\ and\ \bibinfo {author} {\bibfnamefont {U.~L.}\ \bibnamefont {Andersen}},\ }\bibfield  {title} {\bibinfo {title} {Fault-tolerant continuous-variable measurement-based quantum computation architecture},\ }\href {https://doi.org/10.1103/PRXQuantum.2.030325} {\bibfield  {journal} {\bibinfo  {journal} {PRX Quantum}\ }\textbf {\bibinfo {volume} {2}},\ \bibinfo {pages} {030325} (\bibinfo {year} {2021})}\BibitemShut {NoStop}%
\bibitem [{\citenamefont {Bourassa}\ \emph {et~al.}(2021)\citenamefont {Bourassa}, \citenamefont {Alexander}, \citenamefont {Vasmer}, \citenamefont {Patil}, \citenamefont {Tzitrin}, \citenamefont {Matsuura}, \citenamefont {Su}, \citenamefont {Baragiola}, \citenamefont {Guha}, \citenamefont {Dauphinais}, \citenamefont {Sabapathy}, \citenamefont {Menicucci},\ and\ \citenamefont {Dhand}}]{Bourassa_2021}%
  \BibitemOpen
  \bibfield  {author} {\bibinfo {author} {\bibfnamefont {J.~E.}\ \bibnamefont {Bourassa}}, \bibinfo {author} {\bibfnamefont {R.~N.}\ \bibnamefont {Alexander}}, \bibinfo {author} {\bibfnamefont {M.}~\bibnamefont {Vasmer}}, \bibinfo {author} {\bibfnamefont {A.}~\bibnamefont {Patil}}, \bibinfo {author} {\bibfnamefont {I.}~\bibnamefont {Tzitrin}}, \bibinfo {author} {\bibfnamefont {T.}~\bibnamefont {Matsuura}}, \bibinfo {author} {\bibfnamefont {D.}~\bibnamefont {Su}}, \bibinfo {author} {\bibfnamefont {B.~Q.}\ \bibnamefont {Baragiola}}, \bibinfo {author} {\bibfnamefont {S.}~\bibnamefont {Guha}}, \bibinfo {author} {\bibfnamefont {G.}~\bibnamefont {Dauphinais}}, \bibinfo {author} {\bibfnamefont {K.~K.}\ \bibnamefont {Sabapathy}}, \bibinfo {author} {\bibfnamefont {N.~C.}\ \bibnamefont {Menicucci}},\ and\ \bibinfo {author} {\bibfnamefont {I.}~\bibnamefont {Dhand}},\ }\bibfield  {title} {\bibinfo {title} {Blueprint for a {S}calable {P}hotonic {F}ault-{T}olerant {Q}uantum {C}omputer},\ }\href
  {https://doi.org/10.22331/q-2021-02-04-392} {\bibfield  {journal} {\bibinfo  {journal} {{Quantum}}\ }\textbf {\bibinfo {volume} {5}},\ \bibinfo {pages} {392} (\bibinfo {year} {2021})}\BibitemShut {NoStop}%
\bibitem [{\citenamefont {Tzitrin}\ \emph {et~al.}(2021)\citenamefont {Tzitrin}, \citenamefont {Matsuura}, \citenamefont {Alexander}, \citenamefont {Dauphinais}, \citenamefont {Bourassa}, \citenamefont {Sabapathy}, \citenamefont {Menicucci},\ and\ \citenamefont {Dhand}}]{Tzitrin_2021}%
  \BibitemOpen
  \bibfield  {author} {\bibinfo {author} {\bibfnamefont {I.}~\bibnamefont {Tzitrin}}, \bibinfo {author} {\bibfnamefont {T.}~\bibnamefont {Matsuura}}, \bibinfo {author} {\bibfnamefont {R.~N.}\ \bibnamefont {Alexander}}, \bibinfo {author} {\bibfnamefont {G.}~\bibnamefont {Dauphinais}}, \bibinfo {author} {\bibfnamefont {J.~E.}\ \bibnamefont {Bourassa}}, \bibinfo {author} {\bibfnamefont {K.~K.}\ \bibnamefont {Sabapathy}}, \bibinfo {author} {\bibfnamefont {N.~C.}\ \bibnamefont {Menicucci}},\ and\ \bibinfo {author} {\bibfnamefont {I.}~\bibnamefont {Dhand}},\ }\bibfield  {title} {\bibinfo {title} {Fault-tolerant quantum computation with static linear optics},\ }\href {https://doi.org/10.1103/PRXQuantum.2.040353} {\bibfield  {journal} {\bibinfo  {journal} {PRX Quantum}\ }\textbf {\bibinfo {volume} {2}},\ \bibinfo {pages} {040353} (\bibinfo {year} {2021})}\BibitemShut {NoStop}%
\bibitem [{\citenamefont {Hastrup}\ and\ \citenamefont {Andersen}(2022)}]{Hastrup_2022}%
  \BibitemOpen
  \bibfield  {author} {\bibinfo {author} {\bibfnamefont {J.}~\bibnamefont {Hastrup}}\ and\ \bibinfo {author} {\bibfnamefont {U.~L.}\ \bibnamefont {Andersen}},\ }\bibfield  {title} {\bibinfo {title} {All-optical cat-code quantum error correction},\ }\href {https://doi.org/10.1103/PhysRevResearch.4.043065} {\bibfield  {journal} {\bibinfo  {journal} {Phys. Rev. Res.}\ }\textbf {\bibinfo {volume} {4}},\ \bibinfo {pages} {043065} (\bibinfo {year} {2022})}\BibitemShut {NoStop}%
\bibitem [{\citenamefont {Gottesman}\ \emph {et~al.}(2001)\citenamefont {Gottesman}, \citenamefont {Kitaev},\ and\ \citenamefont {Preskill}}]{Gottesman_2001}%
  \BibitemOpen
  \bibfield  {author} {\bibinfo {author} {\bibfnamefont {D.}~\bibnamefont {Gottesman}}, \bibinfo {author} {\bibfnamefont {A.}~\bibnamefont {Kitaev}},\ and\ \bibinfo {author} {\bibfnamefont {J.}~\bibnamefont {Preskill}},\ }\bibfield  {title} {\bibinfo {title} {Encoding a qubit in an oscillator},\ }\bibfield  {journal} {\bibinfo  {journal} {Physical Review A}\ }\textbf {\bibinfo {volume} {64}},\ \href {https://doi.org/10.1103/physreva.64.012310} {10.1103/physreva.64.012310} (\bibinfo {year} {2001})\BibitemShut {NoStop}%
\bibitem [{\citenamefont {Albert}\ \emph {et~al.}(2018)\citenamefont {Albert}, \citenamefont {Noh}, \citenamefont {Duivenvoorden}, \citenamefont {Young}, \citenamefont {Brierley}, \citenamefont {Reinhold}, \citenamefont {Vuillot}, \citenamefont {Li}, \citenamefont {Shen}, \citenamefont {Girvin}, \citenamefont {Terhal},\ and\ \citenamefont {Jiang}}]{Albert}%
  \BibitemOpen
  \bibfield  {author} {\bibinfo {author} {\bibfnamefont {V.~V.}\ \bibnamefont {Albert}}, \bibinfo {author} {\bibfnamefont {K.}~\bibnamefont {Noh}}, \bibinfo {author} {\bibfnamefont {K.}~\bibnamefont {Duivenvoorden}}, \bibinfo {author} {\bibfnamefont {D.~J.}\ \bibnamefont {Young}}, \bibinfo {author} {\bibfnamefont {R.~T.}\ \bibnamefont {Brierley}}, \bibinfo {author} {\bibfnamefont {P.}~\bibnamefont {Reinhold}}, \bibinfo {author} {\bibfnamefont {C.}~\bibnamefont {Vuillot}}, \bibinfo {author} {\bibfnamefont {L.}~\bibnamefont {Li}}, \bibinfo {author} {\bibfnamefont {C.}~\bibnamefont {Shen}}, \bibinfo {author} {\bibfnamefont {S.~M.}\ \bibnamefont {Girvin}}, \bibinfo {author} {\bibfnamefont {B.~M.}\ \bibnamefont {Terhal}},\ and\ \bibinfo {author} {\bibfnamefont {L.}~\bibnamefont {Jiang}},\ }\bibfield  {title} {\bibinfo {title} {Performance and structure of single-mode bosonic codes},\ }\bibfield  {journal} {\bibinfo  {journal} {Physical Review A}\ }\textbf {\bibinfo {volume} {97}},\ \href
  {https://doi.org/10.1103/physreva.97.032346} {10.1103/physreva.97.032346} (\bibinfo {year} {2018})\BibitemShut {NoStop}%
\bibitem [{\citenamefont {Michael}\ \emph {et~al.}(2016)\citenamefont {Michael}, \citenamefont {Silveri}, \citenamefont {Brierley}, \citenamefont {Albert}, \citenamefont {Salmilehto}, \citenamefont {Jiang},\ and\ \citenamefont {Girvin}}]{Michael}%
  \BibitemOpen
  \bibfield  {author} {\bibinfo {author} {\bibfnamefont {M.~H.}\ \bibnamefont {Michael}}, \bibinfo {author} {\bibfnamefont {M.}~\bibnamefont {Silveri}}, \bibinfo {author} {\bibfnamefont {R.~T.}\ \bibnamefont {Brierley}}, \bibinfo {author} {\bibfnamefont {V.~V.}\ \bibnamefont {Albert}}, \bibinfo {author} {\bibfnamefont {J.}~\bibnamefont {Salmilehto}}, \bibinfo {author} {\bibfnamefont {L.}~\bibnamefont {Jiang}},\ and\ \bibinfo {author} {\bibfnamefont {S.~M.}\ \bibnamefont {Girvin}},\ }\bibfield  {title} {\bibinfo {title} {New class of quantum error-correcting codes for a bosonic mode},\ }\href {https://doi.org/10.1103/PhysRevX.6.031006} {\bibfield  {journal} {\bibinfo  {journal} {Phys. Rev. X}\ }\textbf {\bibinfo {volume} {6}},\ \bibinfo {pages} {031006} (\bibinfo {year} {2016})}\BibitemShut {NoStop}%
\bibitem [{\citenamefont {Campagne-Ibarcq}\ \emph {et~al.}(2020)\citenamefont {Campagne-Ibarcq}, \citenamefont {Eickbusch}, \citenamefont {Touzard}, \citenamefont {Zalys-Geller}, \citenamefont {Frattini}, \citenamefont {Sivak}, \citenamefont {Reinhold}, \citenamefont {Puri}, \citenamefont {Shankar}, \citenamefont {Schoelkopf}, \citenamefont {Frunzio}, \citenamefont {Mirrahimi},\ and\ \citenamefont {Devoret}}]{Campagne_Ibarcq_2020}%
  \BibitemOpen
  \bibfield  {author} {\bibinfo {author} {\bibfnamefont {P.}~\bibnamefont {Campagne-Ibarcq}}, \bibinfo {author} {\bibfnamefont {A.}~\bibnamefont {Eickbusch}}, \bibinfo {author} {\bibfnamefont {S.}~\bibnamefont {Touzard}}, \bibinfo {author} {\bibfnamefont {E.}~\bibnamefont {Zalys-Geller}}, \bibinfo {author} {\bibfnamefont {N.~E.}\ \bibnamefont {Frattini}}, \bibinfo {author} {\bibfnamefont {V.~V.}\ \bibnamefont {Sivak}}, \bibinfo {author} {\bibfnamefont {P.}~\bibnamefont {Reinhold}}, \bibinfo {author} {\bibfnamefont {S.}~\bibnamefont {Puri}}, \bibinfo {author} {\bibfnamefont {S.}~\bibnamefont {Shankar}}, \bibinfo {author} {\bibfnamefont {R.~J.}\ \bibnamefont {Schoelkopf}}, \bibinfo {author} {\bibfnamefont {L.}~\bibnamefont {Frunzio}}, \bibinfo {author} {\bibfnamefont {M.}~\bibnamefont {Mirrahimi}},\ and\ \bibinfo {author} {\bibfnamefont {M.~H.}\ \bibnamefont {Devoret}},\ }\bibfield  {title} {\bibinfo {title} {Quantum error correction of a qubit encoded in grid states of an oscillator},\ }\href
  {https://doi.org/10.1038/s41586-020-2603-3} {\bibfield  {journal} {\bibinfo  {journal} {Nature}\ }\textbf {\bibinfo {volume} {584}},\ \bibinfo {pages} {368–372} (\bibinfo {year} {2020})}\BibitemShut {NoStop}%
\bibitem [{\citenamefont {Eickbusch}\ \emph {et~al.}(2021)\citenamefont {Eickbusch}, \citenamefont {Sivak}, \citenamefont {Ding}, \citenamefont {Elder}, \citenamefont {Jha}, \citenamefont {Venkatraman}, \citenamefont {Royer}, \citenamefont {Girvin}, \citenamefont {Schoelkopf},\ and\ \citenamefont {Devoret}}]{Eickbusch-2021}%
  \BibitemOpen
  \bibfield  {author} {\bibinfo {author} {\bibfnamefont {A.}~\bibnamefont {Eickbusch}}, \bibinfo {author} {\bibfnamefont {V.}~\bibnamefont {Sivak}}, \bibinfo {author} {\bibfnamefont {A.~Z.}\ \bibnamefont {Ding}}, \bibinfo {author} {\bibfnamefont {S.~S.}\ \bibnamefont {Elder}}, \bibinfo {author} {\bibfnamefont {S.~R.}\ \bibnamefont {Jha}}, \bibinfo {author} {\bibfnamefont {J.}~\bibnamefont {Venkatraman}}, \bibinfo {author} {\bibfnamefont {B.}~\bibnamefont {Royer}}, \bibinfo {author} {\bibfnamefont {S.}~\bibnamefont {Girvin}}, \bibinfo {author} {\bibfnamefont {R.}~\bibnamefont {Schoelkopf}},\ and\ \bibinfo {author} {\bibfnamefont {M.}~\bibnamefont {Devoret}},\ }\href {https://doi.org/10.1038/s41567-022-01776-9} {\bibinfo {title} {Fast universal control of an oscillator with weak dispersive coupling to a qubit}} (\bibinfo {year} {2021})\BibitemShut {NoStop}%
\bibitem [{\citenamefont {Heeres}\ \emph {et~al.}(2016)\citenamefont {Heeres}, \citenamefont {Reinhold}, \citenamefont {Ofek}, \citenamefont {Frunzio}, \citenamefont {Jiang}, \citenamefont {Devoret},\ and\ \citenamefont {Schoelkopf}}]{Heeres-2016}%
  \BibitemOpen
  \bibfield  {author} {\bibinfo {author} {\bibfnamefont {R.}~\bibnamefont {Heeres}}, \bibinfo {author} {\bibfnamefont {P.}~\bibnamefont {Reinhold}}, \bibinfo {author} {\bibfnamefont {N.}~\bibnamefont {Ofek}}, \bibinfo {author} {\bibfnamefont {L.}~\bibnamefont {Frunzio}}, \bibinfo {author} {\bibfnamefont {L.}~\bibnamefont {Jiang}}, \bibinfo {author} {\bibfnamefont {M.}~\bibnamefont {Devoret}},\ and\ \bibinfo {author} {\bibfnamefont {R.}~\bibnamefont {Schoelkopf}},\ }\href {https://doi.org/10.1038/s41467-017-00045-1} {\bibinfo {title} {Implementing a universal gate set on a logical qubit encoded in an oscillator}} (\bibinfo {year} {2016})\BibitemShut {NoStop}%
\bibitem [{\citenamefont {Didier}\ \emph {et~al.}(2015)\citenamefont {Didier}, \citenamefont {Bourassa},\ and\ \citenamefont {Blais}}]{Didier_2015}%
  \BibitemOpen
  \bibfield  {author} {\bibinfo {author} {\bibfnamefont {N.}~\bibnamefont {Didier}}, \bibinfo {author} {\bibfnamefont {J.}~\bibnamefont {Bourassa}},\ and\ \bibinfo {author} {\bibfnamefont {A.}~\bibnamefont {Blais}},\ }\bibfield  {title} {\bibinfo {title} {Fast quantum nondemolition readout by parametric modulation of longitudinal qubit-oscillator interaction},\ }\bibfield  {journal} {\bibinfo  {journal} {Physical Review Letters}\ }\textbf {\bibinfo {volume} {115}},\ \href {https://doi.org/10.1103/physrevlett.115.203601} {10.1103/physrevlett.115.203601} (\bibinfo {year} {2015})\BibitemShut {NoStop}%
\bibitem [{\citenamefont {Hu}\ \emph {et~al.}(2019)\citenamefont {Hu}, \citenamefont {Ma}, \citenamefont {Cai}, \citenamefont {Mu}, \citenamefont {Xu}, \citenamefont {Wang}, \citenamefont {Wu}, \citenamefont {Wang}, \citenamefont {Song}, \citenamefont {Zou}, \citenamefont {Girvin}, \citenamefont {Duan},\ and\ \citenamefont {Sun}}]{Hu_2019}%
  \BibitemOpen
  \bibfield  {author} {\bibinfo {author} {\bibfnamefont {L.}~\bibnamefont {Hu}}, \bibinfo {author} {\bibfnamefont {Y.}~\bibnamefont {Ma}}, \bibinfo {author} {\bibfnamefont {W.}~\bibnamefont {Cai}}, \bibinfo {author} {\bibfnamefont {X.}~\bibnamefont {Mu}}, \bibinfo {author} {\bibfnamefont {Y.}~\bibnamefont {Xu}}, \bibinfo {author} {\bibfnamefont {W.}~\bibnamefont {Wang}}, \bibinfo {author} {\bibfnamefont {Y.}~\bibnamefont {Wu}}, \bibinfo {author} {\bibfnamefont {H.}~\bibnamefont {Wang}}, \bibinfo {author} {\bibfnamefont {Y.~P.}\ \bibnamefont {Song}}, \bibinfo {author} {\bibfnamefont {C.-L.}\ \bibnamefont {Zou}}, \bibinfo {author} {\bibfnamefont {S.~M.}\ \bibnamefont {Girvin}}, \bibinfo {author} {\bibfnamefont {L.-M.}\ \bibnamefont {Duan}},\ and\ \bibinfo {author} {\bibfnamefont {L.}~\bibnamefont {Sun}},\ }\bibfield  {title} {\bibinfo {title} {Quantum error correction and universal gate set operation on a binomial bosonic logical qubit},\ }\href {https://doi.org/10.1038/s41567-018-0414-3} {\bibfield  {journal}
  {\bibinfo  {journal} {Nature Physics}\ }\textbf {\bibinfo {volume} {15}},\ \bibinfo {pages} {503–508} (\bibinfo {year} {2019})}\BibitemShut {NoStop}%
\bibitem [{\citenamefont {Rosenblum}\ \emph {et~al.}(2018)\citenamefont {Rosenblum}, \citenamefont {Reinhold}, \citenamefont {Mirrahimi}, \citenamefont {Jiang}, \citenamefont {Frunzio},\ and\ \citenamefont {Schoelkopf}}]{Rosenblum_2018}%
  \BibitemOpen
  \bibfield  {author} {\bibinfo {author} {\bibfnamefont {S.}~\bibnamefont {Rosenblum}}, \bibinfo {author} {\bibfnamefont {P.}~\bibnamefont {Reinhold}}, \bibinfo {author} {\bibfnamefont {M.}~\bibnamefont {Mirrahimi}}, \bibinfo {author} {\bibfnamefont {L.}~\bibnamefont {Jiang}}, \bibinfo {author} {\bibfnamefont {L.}~\bibnamefont {Frunzio}},\ and\ \bibinfo {author} {\bibfnamefont {R.~J.}\ \bibnamefont {Schoelkopf}},\ }\bibfield  {title} {\bibinfo {title} {Fault-tolerant detection of a quantum error},\ }\href {https://doi.org/10.1126/science.aat3996} {\bibfield  {journal} {\bibinfo  {journal} {Science}\ }\textbf {\bibinfo {volume} {361}},\ \bibinfo {pages} {266} (\bibinfo {year} {2018})}\BibitemShut {NoStop}%
\bibitem [{\citenamefont {Singh}\ \emph {et~al.}(2025)\citenamefont {Singh}, \citenamefont {Royer},\ and\ \citenamefont {Girvin}}]{Singh_2025}%
  \BibitemOpen
  \bibfield  {author} {\bibinfo {author} {\bibfnamefont {S.}~\bibnamefont {Singh}}, \bibinfo {author} {\bibfnamefont {B.}~\bibnamefont {Royer}},\ and\ \bibinfo {author} {\bibfnamefont {S.~M.}\ \bibnamefont {Girvin}},\ }\href {https://arxiv.org/abs/2504.19992} {\bibinfo {title} {Towards non-abelian quantum signal processing: Efficient control of hybrid continuous- and discrete-variable architectures}} (\bibinfo {year} {2025}),\ \Eprint {https://arxiv.org/abs/2504.19992} {arXiv:2504.19992 [quant-ph]} \BibitemShut {NoStop}%
\bibitem [{\citenamefont {Hastrup}\ \emph {et~al.}(2021)\citenamefont {Hastrup}, \citenamefont {Larsen}, \citenamefont {Neergaard-Nielsen}, \citenamefont {Menicucci},\ and\ \citenamefont {Andersen}}]{Hastrup_2021}%
  \BibitemOpen
  \bibfield  {author} {\bibinfo {author} {\bibfnamefont {J.}~\bibnamefont {Hastrup}}, \bibinfo {author} {\bibfnamefont {M.~V.}\ \bibnamefont {Larsen}}, \bibinfo {author} {\bibfnamefont {J.~S.}\ \bibnamefont {Neergaard-Nielsen}}, \bibinfo {author} {\bibfnamefont {N.~C.}\ \bibnamefont {Menicucci}},\ and\ \bibinfo {author} {\bibfnamefont {U.~L.}\ \bibnamefont {Andersen}},\ }\bibfield  {title} {\bibinfo {title} {Unsuitability of cubic phase gates for non-clifford operations on gottesman-kitaev-preskill states},\ }\href {https://doi.org/10.1103/PhysRevA.103.032409} {\bibfield  {journal} {\bibinfo  {journal} {Phys. Rev. A}\ }\textbf {\bibinfo {volume} {103}},\ \bibinfo {pages} {032409} (\bibinfo {year} {2021})}\BibitemShut {NoStop}%
\bibitem [{\citenamefont {Baragiola}\ \emph {et~al.}(2019)\citenamefont {Baragiola}, \citenamefont {Pantaleoni}, \citenamefont {Alexander}, \citenamefont {Karanjai},\ and\ \citenamefont {Menicucci}}]{Baragiola_2019}%
  \BibitemOpen
  \bibfield  {author} {\bibinfo {author} {\bibfnamefont {B.~Q.}\ \bibnamefont {Baragiola}}, \bibinfo {author} {\bibfnamefont {G.}~\bibnamefont {Pantaleoni}}, \bibinfo {author} {\bibfnamefont {R.~N.}\ \bibnamefont {Alexander}}, \bibinfo {author} {\bibfnamefont {A.}~\bibnamefont {Karanjai}},\ and\ \bibinfo {author} {\bibfnamefont {N.~C.}\ \bibnamefont {Menicucci}},\ }\bibfield  {title} {\bibinfo {title} {All-gaussian universality and fault tolerance with the gottesman-kitaev-preskill code},\ }\bibfield  {journal} {\bibinfo  {journal} {Physical Review Letters}\ }\textbf {\bibinfo {volume} {123}},\ \href {https://doi.org/10.1103/physrevlett.123.200502} {10.1103/physrevlett.123.200502} (\bibinfo {year} {2019})\BibitemShut {NoStop}%
\bibitem [{\citenamefont {Ding}(2025)}]{Ding_2025}%
  \BibitemOpen
  \bibfield  {author} {\bibinfo {author} {\bibfnamefont {B.~B. E. A. e.~a.}\ \bibnamefont {Ding}, \bibfnamefont {A.Z.}},\ }\bibfield  {title} {\bibinfo {title} {Quantum control of an oscillator with a kerr-cat qubit},\ }\bibfield  {journal} {\bibinfo  {journal} {Nat Commun}\ }\href {https://doi.org/10.1038/s41586-020-2587-z} {10.1038/s41586-020-2587-z} (\bibinfo {year} {2025})\BibitemShut {NoStop}%
\bibitem [{\citenamefont {Royer}\ \emph {et~al.}(2020)\citenamefont {Royer}, \citenamefont {Singh},\ and\ \citenamefont {Girvin}}]{Royer_2020}%
  \BibitemOpen
  \bibfield  {author} {\bibinfo {author} {\bibfnamefont {B.}~\bibnamefont {Royer}}, \bibinfo {author} {\bibfnamefont {S.}~\bibnamefont {Singh}},\ and\ \bibinfo {author} {\bibfnamefont {S.~M.}\ \bibnamefont {Girvin}},\ }\bibfield  {title} {\bibinfo {title} {Stabilization of finite-energy gottesman-kitaev-preskill states},\ }\href {https://doi.org/10.1103/PhysRevLett.125.260509} {\bibfield  {journal} {\bibinfo  {journal} {Phys. Rev. Lett.}\ }\textbf {\bibinfo {volume} {125}},\ \bibinfo {pages} {260509} (\bibinfo {year} {2020})}\BibitemShut {NoStop}%
\bibitem [{\citenamefont {Hopfmueller}\ \emph {et~al.}(2024)\citenamefont {Hopfmueller}, \citenamefont {Tremblay}, \citenamefont {St-Jean}, \citenamefont {Royer},\ and\ \citenamefont {Lemonde}}]{Hopfmueller_2024}%
  \BibitemOpen
  \bibfield  {author} {\bibinfo {author} {\bibfnamefont {F.}~\bibnamefont {Hopfmueller}}, \bibinfo {author} {\bibfnamefont {M.}~\bibnamefont {Tremblay}}, \bibinfo {author} {\bibfnamefont {P.}~\bibnamefont {St-Jean}}, \bibinfo {author} {\bibfnamefont {B.}~\bibnamefont {Royer}},\ and\ \bibinfo {author} {\bibfnamefont {M.-A.}\ \bibnamefont {Lemonde}},\ }\bibfield  {title} {\bibinfo {title} {Bosonic pauli+: Efficient simulation of concatenated gottesman-kitaev-preskill codes},\ }\href {https://doi.org/10.22331/q-2024-11-26-1539} {\bibfield  {journal} {\bibinfo  {journal} {Quantum}\ }\textbf {\bibinfo {volume} {8}},\ \bibinfo {pages} {1539} (\bibinfo {year} {2024})}\BibitemShut {NoStop}%
\bibitem [{Note1()}]{Note1}%
  \BibitemOpen
  \bibinfo {note} {Note that $\protect \hat P_\Delta $ is not quite a projector since the code words defined by \protect \cref {eq:finiteEnergyStates} are not perfectly orthogonal.}\BibitemShut {Stop}%
\bibitem [{\citenamefont {Bravyi}\ and\ \citenamefont {Kitaev}(2005)}]{Bravyi_2005}%
  \BibitemOpen
  \bibfield  {author} {\bibinfo {author} {\bibfnamefont {S.}~\bibnamefont {Bravyi}}\ and\ \bibinfo {author} {\bibfnamefont {A.}~\bibnamefont {Kitaev}},\ }\bibfield  {title} {\bibinfo {title} {Universal quantum computation with ideal clifford gates and noisy ancillas},\ }\bibfield  {journal} {\bibinfo  {journal} {Physical Review A}\ }\textbf {\bibinfo {volume} {71}},\ \href {https://doi.org/10.1103/physreva.71.022316} {10.1103/physreva.71.022316} (\bibinfo {year} {2005})\BibitemShut {NoStop}%
\bibitem [{\citenamefont {Blais}\ \emph {et~al.}(2021)\citenamefont {Blais}, \citenamefont {Grimsmo}, \citenamefont {Girvin},\ and\ \citenamefont {Wallraff}}]{Alexandre_RMP}%
  \BibitemOpen
  \bibfield  {author} {\bibinfo {author} {\bibfnamefont {A.}~\bibnamefont {Blais}}, \bibinfo {author} {\bibfnamefont {A.~L.}\ \bibnamefont {Grimsmo}}, \bibinfo {author} {\bibfnamefont {S.~M.}\ \bibnamefont {Girvin}},\ and\ \bibinfo {author} {\bibfnamefont {A.}~\bibnamefont {Wallraff}},\ }\bibfield  {title} {\bibinfo {title} {Circuit quantum electrodynamics},\ }\href {https://doi.org/10.1103/RevModPhys.93.025005} {\bibfield  {journal} {\bibinfo  {journal} {Rev. Mod. Phys.}\ }\textbf {\bibinfo {volume} {93}},\ \bibinfo {pages} {025005} (\bibinfo {year} {2021})}\BibitemShut {NoStop}%
\bibitem [{\citenamefont {Hua}\ \emph {et~al.}(2025)\citenamefont {Hua}, \citenamefont {Xu}, \citenamefont {Wang}, \citenamefont {Ma}, \citenamefont {Zhou}, \citenamefont {Cai}, \citenamefont {Ai}, \citenamefont {Liu}, \citenamefont {Li}, \citenamefont {Zou},\ and\ \citenamefont {Sun}}]{Hua_2025}%
  \BibitemOpen
  \bibfield  {author} {\bibinfo {author} {\bibfnamefont {Z.}~\bibnamefont {Hua}}, \bibinfo {author} {\bibfnamefont {Y.}~\bibnamefont {Xu}}, \bibinfo {author} {\bibfnamefont {W.}~\bibnamefont {Wang}}, \bibinfo {author} {\bibfnamefont {Y.}~\bibnamefont {Ma}}, \bibinfo {author} {\bibfnamefont {J.}~\bibnamefont {Zhou}}, \bibinfo {author} {\bibfnamefont {W.}~\bibnamefont {Cai}}, \bibinfo {author} {\bibfnamefont {H.}~\bibnamefont {Ai}}, \bibinfo {author} {\bibfnamefont {Y.-x.}\ \bibnamefont {Liu}}, \bibinfo {author} {\bibfnamefont {M.}~\bibnamefont {Li}}, \bibinfo {author} {\bibfnamefont {C.-L.}\ \bibnamefont {Zou}},\ and\ \bibinfo {author} {\bibfnamefont {L.}~\bibnamefont {Sun}},\ }\bibfield  {title} {\bibinfo {title} {Engineering the nonlinearity of bosonic modes with a multiloop squid},\ }\href {https://doi.org/10.1103/PhysRevApplied.23.054031} {\bibfield  {journal} {\bibinfo  {journal} {Phys. Rev. Appl.}\ }\textbf {\bibinfo {volume} {23}},\ \bibinfo {pages} {054031} (\bibinfo {year} {2025})}\BibitemShut
  {NoStop}%
\bibitem [{\citenamefont {Zhang}\ \emph {et~al.}(2017)\citenamefont {Zhang}, \citenamefont {Huang}, \citenamefont {Gershenson},\ and\ \citenamefont {Bell}}]{Zhang_2017}%
  \BibitemOpen
  \bibfield  {author} {\bibinfo {author} {\bibfnamefont {W.}~\bibnamefont {Zhang}}, \bibinfo {author} {\bibfnamefont {W.}~\bibnamefont {Huang}}, \bibinfo {author} {\bibfnamefont {M.~E.}\ \bibnamefont {Gershenson}},\ and\ \bibinfo {author} {\bibfnamefont {M.~T.}\ \bibnamefont {Bell}},\ }\bibfield  {title} {\bibinfo {title} {Josephson metamaterial with a widely tunable positive or negative kerr constant},\ }\href {https://doi.org/10.1103/PhysRevApplied.8.051001} {\bibfield  {journal} {\bibinfo  {journal} {Phys. Rev. Appl.}\ }\textbf {\bibinfo {volume} {8}},\ \bibinfo {pages} {051001} (\bibinfo {year} {2017})}\BibitemShut {NoStop}%
\bibitem [{\citenamefont {Ye}\ \emph {et~al.}(2021)\citenamefont {Ye}, \citenamefont {Peng}, \citenamefont {Naghiloo}, \citenamefont {Cunningham},\ and\ \citenamefont {O'Brien}}]{Ye_2021}%
  \BibitemOpen
  \bibfield  {author} {\bibinfo {author} {\bibfnamefont {Y.}~\bibnamefont {Ye}}, \bibinfo {author} {\bibfnamefont {K.}~\bibnamefont {Peng}}, \bibinfo {author} {\bibfnamefont {M.}~\bibnamefont {Naghiloo}}, \bibinfo {author} {\bibfnamefont {G.}~\bibnamefont {Cunningham}},\ and\ \bibinfo {author} {\bibfnamefont {K.~P.}\ \bibnamefont {O'Brien}},\ }\bibfield  {title} {\bibinfo {title} {Engineering purely nonlinear coupling between superconducting qubits using a quarton},\ }\href {https://doi.org/10.1103/PhysRevLett.127.050502} {\bibfield  {journal} {\bibinfo  {journal} {Phys. Rev. Lett.}\ }\textbf {\bibinfo {volume} {127}},\ \bibinfo {pages} {050502} (\bibinfo {year} {2021})}\BibitemShut {NoStop}%
\bibitem [{\citenamefont {Ye}\ \emph {et~al.}(2024)\citenamefont {Ye}, \citenamefont {Kline}, \citenamefont {Yen}, \citenamefont {Cunningham}, \citenamefont {Tan}, \citenamefont {Zang}, \citenamefont {Gingras}, \citenamefont {Niedzielski}, \citenamefont {Stickler}, \citenamefont {Serniak}, \citenamefont {Schwartz},\ and\ \citenamefont {O'Brien}}]{Ye_2024}%
  \BibitemOpen
  \bibfield  {author} {\bibinfo {author} {\bibfnamefont {Y.}~\bibnamefont {Ye}}, \bibinfo {author} {\bibfnamefont {J.~B.}\ \bibnamefont {Kline}}, \bibinfo {author} {\bibfnamefont {A.}~\bibnamefont {Yen}}, \bibinfo {author} {\bibfnamefont {G.}~\bibnamefont {Cunningham}}, \bibinfo {author} {\bibfnamefont {M.}~\bibnamefont {Tan}}, \bibinfo {author} {\bibfnamefont {A.}~\bibnamefont {Zang}}, \bibinfo {author} {\bibfnamefont {M.}~\bibnamefont {Gingras}}, \bibinfo {author} {\bibfnamefont {B.~M.}\ \bibnamefont {Niedzielski}}, \bibinfo {author} {\bibfnamefont {H.}~\bibnamefont {Stickler}}, \bibinfo {author} {\bibfnamefont {K.}~\bibnamefont {Serniak}}, \bibinfo {author} {\bibfnamefont {M.~E.}\ \bibnamefont {Schwartz}},\ and\ \bibinfo {author} {\bibfnamefont {K.~P.}\ \bibnamefont {O'Brien}},\ }\href {https://arxiv.org/abs/2404.19199} {\bibinfo {title} {Near-ultrastrong nonlinear light-matter coupling in superconducting circuits}} (\bibinfo {year} {2024}),\ \Eprint {https://arxiv.org/abs/2404.19199} {arXiv:2404.19199
  [quant-ph]} \BibitemShut {NoStop}%
\bibitem [{\citenamefont {Elliott}\ \emph {et~al.}(2018)\citenamefont {Elliott}, \citenamefont {Joo},\ and\ \citenamefont {Ginossar}}]{Elliott_2018}%
  \BibitemOpen
  \bibfield  {author} {\bibinfo {author} {\bibfnamefont {M.}~\bibnamefont {Elliott}}, \bibinfo {author} {\bibfnamefont {J.}~\bibnamefont {Joo}},\ and\ \bibinfo {author} {\bibfnamefont {E.}~\bibnamefont {Ginossar}},\ }\bibfield  {title} {\bibinfo {title} {Designing kerr interactions using multiple superconducting qubit types in a single circuit},\ }\href {https://doi.org/10.1088/1367-2630/aa9243} {\bibfield  {journal} {\bibinfo  {journal} {New Journal of Physics}\ }\textbf {\bibinfo {volume} {20}},\ \bibinfo {pages} {023037} (\bibinfo {year} {2018})}\BibitemShut {NoStop}%
\bibitem [{\citenamefont {Scigliuzzo}\ \emph {et~al.}(2025)\citenamefont {Scigliuzzo}, \citenamefont {Peyruchat}, \citenamefont {Marabini}, \citenamefont {Becker}, \citenamefont {Jouanny}, \citenamefont {Delsing},\ and\ \citenamefont {Scarlino}}]{Scigliuzzo_2025}%
  \BibitemOpen
  \bibfield  {author} {\bibinfo {author} {\bibfnamefont {M.}~\bibnamefont {Scigliuzzo}}, \bibinfo {author} {\bibfnamefont {L.}~\bibnamefont {Peyruchat}}, \bibinfo {author} {\bibfnamefont {R.~M.}\ \bibnamefont {Marabini}}, \bibinfo {author} {\bibfnamefont {C.}~\bibnamefont {Becker}}, \bibinfo {author} {\bibfnamefont {V.}~\bibnamefont {Jouanny}}, \bibinfo {author} {\bibfnamefont {P.}~\bibnamefont {Delsing}},\ and\ \bibinfo {author} {\bibfnamefont {P.}~\bibnamefont {Scarlino}},\ }\href {https://arxiv.org/abs/2505.24865} {\bibinfo {title} {Quantum acoustics with tunable nonlinearity in the superstrong coupling regime}} (\bibinfo {year} {2025}),\ \Eprint {https://arxiv.org/abs/2505.24865} {arXiv:2505.24865 [quant-ph]} \BibitemShut {NoStop}%
\bibitem [{\citenamefont {Miano}\ \emph {et~al.}(2022)\citenamefont {Miano}, \citenamefont {Liu}, \citenamefont {Sivak}, \citenamefont {Frattini}, \citenamefont {Joshi}, \citenamefont {Dai}, \citenamefont {Frunzio},\ and\ \citenamefont {Devoret}}]{Miano_2022}%
  \BibitemOpen
  \bibfield  {author} {\bibinfo {author} {\bibfnamefont {A.}~\bibnamefont {Miano}}, \bibinfo {author} {\bibfnamefont {G.}~\bibnamefont {Liu}}, \bibinfo {author} {\bibfnamefont {V.~V.}\ \bibnamefont {Sivak}}, \bibinfo {author} {\bibfnamefont {N.~E.}\ \bibnamefont {Frattini}}, \bibinfo {author} {\bibfnamefont {V.~R.}\ \bibnamefont {Joshi}}, \bibinfo {author} {\bibfnamefont {W.}~\bibnamefont {Dai}}, \bibinfo {author} {\bibfnamefont {L.}~\bibnamefont {Frunzio}},\ and\ \bibinfo {author} {\bibfnamefont {M.~H.}\ \bibnamefont {Devoret}},\ }\bibfield  {title} {\bibinfo {title} {Frequency-tunable kerr-free three-wave mixing with a gradiometric snail},\ }\bibfield  {journal} {\bibinfo  {journal} {Applied Physics Letters}\ }\textbf {\bibinfo {volume} {120}},\ \href {https://doi.org/10.1063/5.0083350} {10.1063/5.0083350} (\bibinfo {year} {2022})\BibitemShut {NoStop}%
\bibitem [{\citenamefont {Sivak}\ \emph {et~al.}(2019)\citenamefont {Sivak}, \citenamefont {Frattini}, \citenamefont {Joshi}, \citenamefont {Lingenfelter}, \citenamefont {Shankar},\ and\ \citenamefont {Devoret}}]{Sivak_2019}%
  \BibitemOpen
  \bibfield  {author} {\bibinfo {author} {\bibfnamefont {V.}~\bibnamefont {Sivak}}, \bibinfo {author} {\bibfnamefont {N.}~\bibnamefont {Frattini}}, \bibinfo {author} {\bibfnamefont {V.}~\bibnamefont {Joshi}}, \bibinfo {author} {\bibfnamefont {A.}~\bibnamefont {Lingenfelter}}, \bibinfo {author} {\bibfnamefont {S.}~\bibnamefont {Shankar}},\ and\ \bibinfo {author} {\bibfnamefont {M.}~\bibnamefont {Devoret}},\ }\bibfield  {title} {\bibinfo {title} {Kerr-free three-wave mixing in superconducting quantum circuits},\ }\href {https://doi.org/10.1103/PhysRevApplied.11.054060} {\bibfield  {journal} {\bibinfo  {journal} {Phys. Rev. Appl.}\ }\textbf {\bibinfo {volume} {11}},\ \bibinfo {pages} {054060} (\bibinfo {year} {2019})}\BibitemShut {NoStop}%
\bibitem [{\citenamefont {He}\ \emph {et~al.}(2023{\natexlab{a}})\citenamefont {He}, \citenamefont {Lu}, \citenamefont {Bao}, \citenamefont {Xue}, \citenamefont {Jiang}, \citenamefont {Wang}, \citenamefont {Roudsari}, \citenamefont {Delsing}, \citenamefont {Tsai},\ and\ \citenamefont {Lin}}]{He_2023}%
  \BibitemOpen
  \bibfield  {author} {\bibinfo {author} {\bibfnamefont {X.~L.}\ \bibnamefont {He}}, \bibinfo {author} {\bibfnamefont {Y.}~\bibnamefont {Lu}}, \bibinfo {author} {\bibfnamefont {D.~Q.}\ \bibnamefont {Bao}}, \bibinfo {author} {\bibfnamefont {H.}~\bibnamefont {Xue}}, \bibinfo {author} {\bibfnamefont {W.~B.}\ \bibnamefont {Jiang}}, \bibinfo {author} {\bibfnamefont {Z.}~\bibnamefont {Wang}}, \bibinfo {author} {\bibfnamefont {A.~F.}\ \bibnamefont {Roudsari}}, \bibinfo {author} {\bibfnamefont {P.}~\bibnamefont {Delsing}}, \bibinfo {author} {\bibfnamefont {J.~S.}\ \bibnamefont {Tsai}},\ and\ \bibinfo {author} {\bibfnamefont {Z.~R.}\ \bibnamefont {Lin}},\ }\href {https://arxiv.org/abs/2308.14676} {\bibinfo {title} {Fast generation of schr\"odinger cat states in a kerr-tunable superconducting resonator}} (\bibinfo {year} {2023}{\natexlab{a}}),\ \Eprint {https://arxiv.org/abs/2308.14676} {arXiv:2308.14676 [quant-ph]} \BibitemShut {NoStop}%
\bibitem [{\citenamefont {Ranadive}\ \emph {et~al.}(2022)\citenamefont {Ranadive}, \citenamefont {Esposito}, \citenamefont {Planat}, \citenamefont {Bonet}, \citenamefont {Naud}, \citenamefont {Buisson}, \citenamefont {Guichard},\ and\ \citenamefont {Roch}}]{Ranadive_2022}%
  \BibitemOpen
  \bibfield  {author} {\bibinfo {author} {\bibfnamefont {A.}~\bibnamefont {Ranadive}}, \bibinfo {author} {\bibfnamefont {M.}~\bibnamefont {Esposito}}, \bibinfo {author} {\bibfnamefont {L.}~\bibnamefont {Planat}}, \bibinfo {author} {\bibfnamefont {E.}~\bibnamefont {Bonet}}, \bibinfo {author} {\bibfnamefont {C.}~\bibnamefont {Naud}}, \bibinfo {author} {\bibfnamefont {O.}~\bibnamefont {Buisson}}, \bibinfo {author} {\bibfnamefont {W.}~\bibnamefont {Guichard}},\ and\ \bibinfo {author} {\bibfnamefont {N.}~\bibnamefont {Roch}},\ }\bibfield  {title} {\bibinfo {title} {Kerr reversal in josephson meta-material and traveling wave parametric amplification},\ }\bibfield  {journal} {\bibinfo  {journal} {Nature Communications}\ }\textbf {\bibinfo {volume} {13}},\ \href {https://doi.org/10.1038/s41467-022-29375-5} {10.1038/s41467-022-29375-5} (\bibinfo {year} {2022})\BibitemShut {NoStop}%
\bibitem [{\citenamefont {He}\ \emph {et~al.}(2023{\natexlab{b}})\citenamefont {He}, \citenamefont {Lu}, \citenamefont {Bao}, \citenamefont {Xue}, \citenamefont {Jiang}, \citenamefont {Wang}, \citenamefont {Roudsari}, \citenamefont {Delsing}, \citenamefont {Tsai},\ and\ \citenamefont {Lin}}]{He2023}%
  \BibitemOpen
  \bibfield  {author} {\bibinfo {author} {\bibfnamefont {X.~L.}\ \bibnamefont {He}}, \bibinfo {author} {\bibfnamefont {Y.}~\bibnamefont {Lu}}, \bibinfo {author} {\bibfnamefont {D.~Q.}\ \bibnamefont {Bao}}, \bibinfo {author} {\bibfnamefont {H.}~\bibnamefont {Xue}}, \bibinfo {author} {\bibfnamefont {W.~B.}\ \bibnamefont {Jiang}}, \bibinfo {author} {\bibfnamefont {Z.}~\bibnamefont {Wang}}, \bibinfo {author} {\bibfnamefont {A.~F.}\ \bibnamefont {Roudsari}}, \bibinfo {author} {\bibfnamefont {P.}~\bibnamefont {Delsing}}, \bibinfo {author} {\bibfnamefont {J.~S.}\ \bibnamefont {Tsai}},\ and\ \bibinfo {author} {\bibfnamefont {Z.~R.}\ \bibnamefont {Lin}},\ }\bibfield  {title} {\bibinfo {title} {Fast generation of schrödinger cat states using a kerr-tunable superconducting resonator},\ }\href {https://doi.org/10.1038/s41467-023-42057-0} {\bibfield  {journal} {\bibinfo  {journal} {Nature Communications}\ }\textbf {\bibinfo {volume} {14}},\ \bibinfo {pages} {6358} (\bibinfo {year} {2023}{\natexlab{b}})}\BibitemShut
  {NoStop}%
\end{thebibliography}%
\clearpage

\appendix



\section*{Supplementary material}

# Supplemental Material for
# "Using a Kerr interaction for GKP magic state preparation"

Jérémie Boudreault,[1] Ross Shillito,[2] Jean-Baptiste Bertrand,[3] and Baptiste Royer[3]

[1] *Département de Physique and Institut Quantique,*
*Université de Sherbrooke, Sherbrooke J1K 2R1 Québec, CAN*
[2] *Nord Quantique, Sherbrooke J1K 2X8 Québec, CAN*
[3] *Départment de Physique and Institut Quantique,*
*Université de Sherbrooke, Sherbrooke J1K 2R1 Québec, CAN*

### CONTENTS

## I. FINITE-ENERGY GKP

### A. States

In this section we derive the Fock state representation of finite-energy GKP states. Starting from an ideal state in the position basis,

$$|\mu_0\rangle \propto \sum_{j \in \mathbb{Z}} |2\sqrt{\pi}(j + \mu/2)\rangle_q,$$

we have defined the finite-energy state as

$$|\mu_\Delta\rangle = \mathcal{N}_{\Delta,\mu} \hat{E}_\Delta \sum_{j \in \mathbb{Z}} |2\sqrt{\pi}(j + \mu/2)\rangle_q, \tag{S1}$$

where $\hat{E}_\Delta = e^{-\Delta^2 \hat{n}}$ is the Gaussian envelope operator and $\mathcal{N}_{\Delta,\mu}$ is a normalisation constant. Switching to the Fock basis,

$$|\mu_\Delta\rangle = \mathcal{N}_{\Delta,\mu} \sum_{n=0}^{\infty} \sum_{j \in \mathbb{Z}} |n\rangle\langle n|\hat{E}_\Delta |2\sqrt{\pi}(j + \mu/2)\rangle_q, \tag{S2}$$

$$= \mathcal{N}_{\Delta,\mu} \sum_{n=0}^{\infty} c_n |n\rangle \tag{S3}$$

with

$$c_n = e^{-\Delta^2 n} \sum_{j \in \mathbb{Z}} \langle n|2\sqrt{\pi}(j + \mu/2)\rangle_q, \tag{S4}$$

$$= e^{-\Delta^2 n} \sum_{j \in \mathbb{Z}} \frac{e^{-(2\sqrt{\pi}(j+\frac{\mu}{2}))^2/2}}{\sqrt{2^n n! \sqrt{\pi}}} \mathcal{H}_n(2\sqrt{\pi}(j + \mu/2)). \tag{S5}$$

Here $\mathcal{H}_n(x)$ is the $n^{th}$ physicist's Hermite polynomial. Due to the symmetry of these polynomials, we can make simplifications. First, for all odd $n$, we have $c_{n_{\text{odd}}} = 0$ because $\mathcal{H}_{n_{\text{odd}}}(x) = -\mathcal{H}_{n_{\text{odd}}}(-x)$. Also, for even terms, we have $\mathcal{H}_{n_{\text{even}}}(x) = \mathcal{H}_{n_{\text{even}}}(-x)$, thus we can simplify the sum over $j$ by keeping only $j > 0$ terms and multiplying by a factor of 2. In the end, we get

$$|0_\Delta\rangle = \mathcal{N}_{\Delta,0} \sum_{n=0}^{n_{\max}} c_{2n}|2n\rangle, \tag{S6}$$

with

$$c_{2n} = \frac{e^{-\Delta^2 2n}}{\sqrt{2^n (2n)! \sqrt{\pi}}} \left[(1 - \mu)\mathcal{H}_{2n}(0) + 2 \sum_{j=1}^{j_{\max}} e^{-(2\sqrt{\pi}(j+\frac{\mu}{2}))^2/2} \mathcal{H}_{2n}(2\sqrt{\pi}(j + \mu/2))\right] \tag{S7}$$

and the normalisation constant is calculated numerically. We truncate the summation at $j_{\max} = 10$, and we set $n_{\max}$ depending on the size of the GKP state under consideration.

### B. Calculation of the $T$ gate envelope distorsion

When preparing a GKP magic state with a cubic gate, the envelope gets non-trivially distorted. In order to get an accurate depiction of it's final shape, we simulate the dynamic of some initial $(x, p)$ phase-space points under the equivalent cubic classical Hamiltonian. As mentioned in [1], an logical $T_L$ gate can be implemented by the operator

$$\hat{U}_T = \exp\left[i\frac{\pi}{4}\left(\frac{2\hat{q}^3}{\pi^{3/2}} + \frac{\hat{q}^2}{\pi} - \frac{2\hat{q}}{\sqrt{\pi}}\right)\right]. \tag{S8}$$

If we only consider the trajectory of a point $(q, p)$ in phase space, it will evolve according to the classical hamilton's equations of motion with the hamiltonian

$$\mathcal{H} = -\frac{2q^3}{\pi^{3/2}} - \frac{q^2}{\pi} + \frac{2q}{\sqrt{\pi}}$$

during a time $t = \pi/4$. Explicitly, we have

$$-\frac{\partial \mathcal{H}}{\partial x} = \frac{6x^2}{\pi^{3/2}} + \frac{2x}{\pi} - \frac{2}{\sqrt{\pi}} \quad \text{and} \quad \frac{\partial \mathcal{H}}{\partial p} = 0.$$

Solving numerically for points of an initial circular envelope of radius $r = 2\sqrt{2\pi}$, we get the shape illustrated in figure 1(d) of the main text.

## II. KERR EVOLUTION IN THE SBS BASIS

The SBS basis introduced in Ref. [2] is a sub-system decomposition of the bosonic Hilbert space, $\mathcal{H}_P$, as a tensor product of a logical qubit space and an error space, $\mathcal{H}_P = \mathcal{H}_L \otimes \mathcal{H}_e$. The error states are labeled $(e_q, e_p)$ and are spanned by states $(\hat{K}_{pe}^\dagger)^{e_p}(\hat{K}_{qe}^\dagger)^{e_q}|\mu_\Delta\rangle$, with $\mu \in \{0, 1\}$ and $\hat{K}_{qe}, \hat{K}_{pe}$ the SBS Kraus operators for correcting an error in the $q$ and $p$ quadrature coordinates, respectively. The subsystem decomposition allow us to extract logical information spread over the entirety of the Hilbert space, i.e. we define the logical sBs qubit as $\rho_L = \text{Tr}_e[\rho_P]$. Figure S1 a) illustrates the action of $\hat{U}_K$ on the logical information contained in all error subspaces. Figure S1 b) highlights the undesired part of the $\hat{U}_K$ with $\hat{U}_K^{err} = (\sqrt{H} \otimes \mathbb{I})^\dagger \hat{U}_K$. The latter indeed applies the identity in the $(0, 0)$ subspace, but corrupts the logical content in every other error subspace.

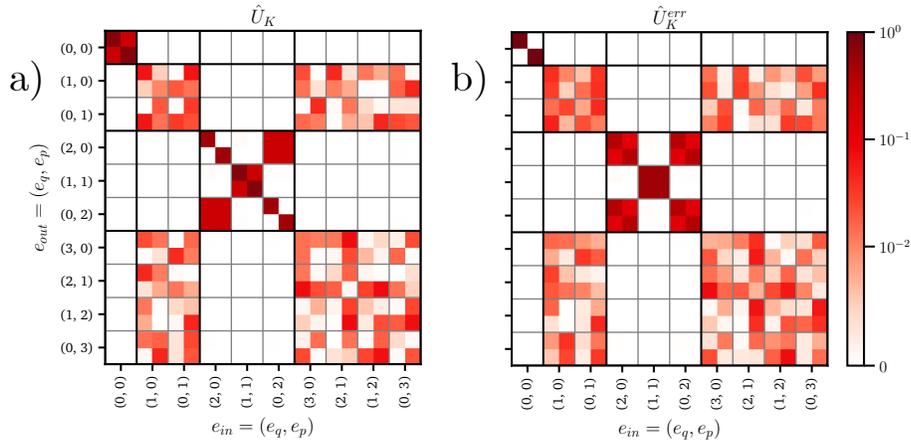

FIG. S1. Amplitudes squared of a) $\hat{U}_K$, b) $\hat{U}_K^{err}$ in the SBS basis.

## III. ERROR ANALYSIS

### A. Dephasing channel

This section outlines the effect of dephasing with the dissipator $\mathcal{D}[\sqrt{\kappa_\phi}\hat{n}]$ on magic state preparation with the Kerr unitary $\hat{U}_K$. When scaling dephasing proportionally with photon loss ($\kappa = \kappa_\phi$), the fidelity is initially limited by loss for ($\gamma < 10^{-3}$), but then is limited by dephasing in the regime ($\gamma \gtrsim 10^{-2}$). The effect of dephasing is similarly amplified for larger GKP states.

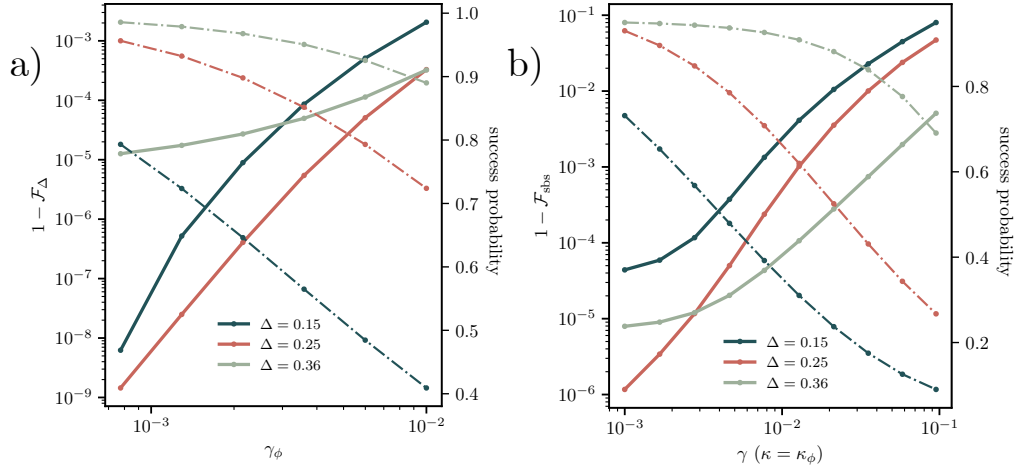

FIG. S2. Magic state infidelity as function of the dephasing parameter $\gamma_\phi = \kappa_\phi t_K$ for a) pure dephasing with perfect error detection, b) dephasing and photon loss with $N = 30$ rounds of successful SBS post-selection. In the latter the loss and dephasing parameters scale at identical rates.

### B. Measurement errors

In the main Letter, we showed that a large number of post-selection rounds enables recovery of the quadratic gain in fidelity as a function of loss. In practice, post-selection protocols are themselves susceptible to measurement errors. Then, increasing post-selection does not guarantee better fidelities, and there is

often an optimal number of rounds before the fidelity begins to deteriorate. In the case when $\mathrm{P}(e|g) \gtrsim \gamma$, it is preferable not to apply any post-selection at all. However, one can minimize the probability $\mathrm{P}(e|g)$ of these errors by adjusting the measurement cut-off in an asymmetric manner, at the cost of increasing the probability $\mathrm{P}(g|e)$[3]. The yield decreases as more valid measurements are discarded, but the quality of the post-selection increases.

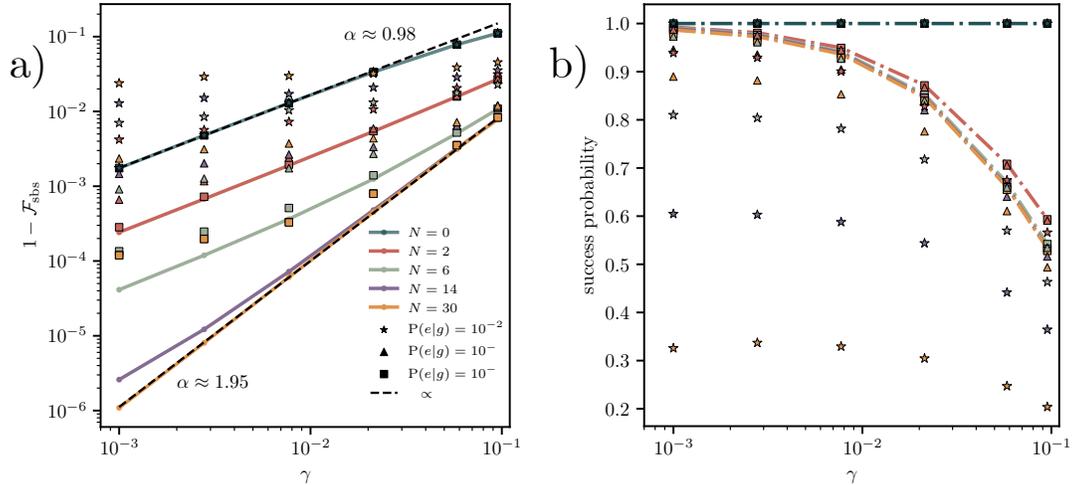

FIG. S3. a) Magic state infidelity and b) the success probability of the post-selection procedure with respect to photon loss for various $N$ number of successful SBS rounds. Solid lines represent ideal post-selection while symbols represent various measurement error probabilities $\mathrm{P}(e|g)$.

### C. Noisy input state

Because the unitary $\hat{U}_K$ is non-Gaussian, it amplifies errors as they occur. It is therefore interesting to consider a noiseless magic state preparation from a noisy input to isolate the effect of initial errors. We therefore consider a noiseless magic state preparation scenario where an ideal Pauli state first undergoes noisy SBS error correction, followed by application of the perfect unitary $\hat{U}_K$. The circuit is shown on figure S4.

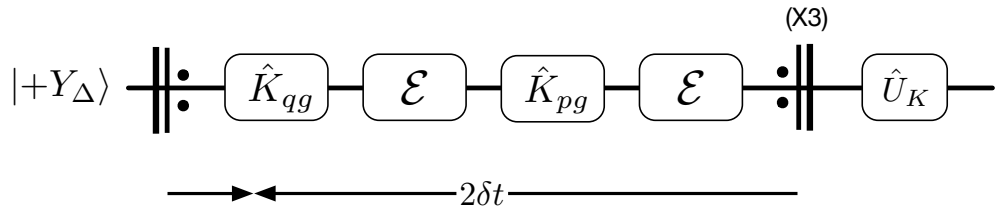

FIG. S4. Magic state preparation from a noisy Pauli state. Starting from an ideal Pauli state, we apply three full rounds of SBS error correction under the loss channel $\mathcal{E}$ expressed by the Kraus operators $\hat{E}_\ell = \left(\frac{\gamma}{1-\gamma}\right)^{\ell/2} \frac{\hat{a}^\ell}{\sqrt{\ell!}} \left(1 - \gamma\right)^{\hat{n}/2}$ that we truncate at $\ell = 10$. We apply post-selection on successful preparations only, as the SBS Kraus operators $\hat{K}_{qg}$ and $\hat{K}_{pg}$ apply correction when no error was detected in the $q$ and $p$ quadrature coordinates respectively. We assume for simplicity that $t_{\hat{U}_K} \approx \delta t$ in our definition of $\gamma$. We also consider that $\hat{U}_K$ is perfect, such that the errors only come from the initial noise.

We observe a transition near $\gamma = 10^{-2}$, beyond which fidelities deteriorate rapidly due to the spreading of logical information into the error subspaces. The probability of measuring an error increases and the logical information degrades.


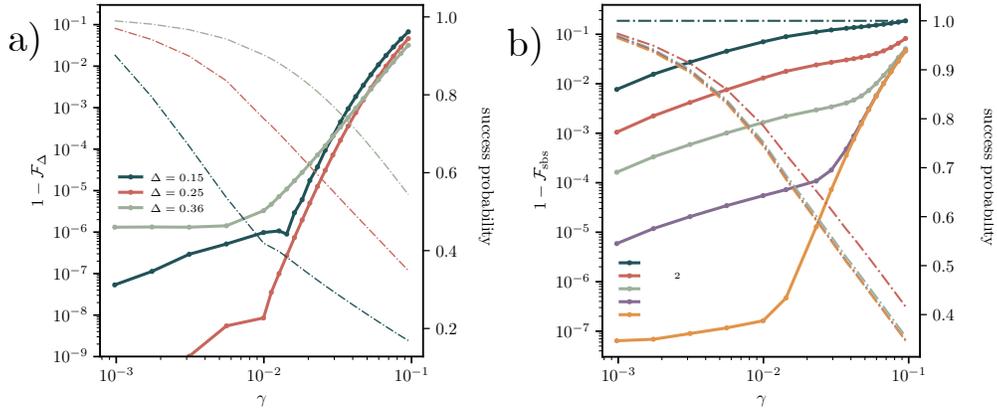

FIG. S5. Infidelities of magic state preparation with a noisy Pauli state and an ideal unitary $\hat{U}_K$, considering a) perfect error detection decoding and b) various rounds of full SBS error correction and post-selection for $\Delta = 0.25$.

## IV. CIRCUIT QED IMPLEMENTATION OF SELF-KERR GATE

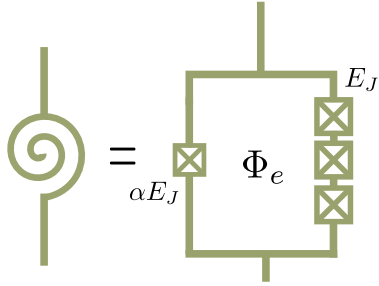

FIG. S6. The SNAIL circuit. Here, we choose n=3 Josephson junctions in the junction array each with Josephson energy $E_J$, and with a SNAIL junction asymmetry $\alpha = 0.07$. By changing the external flux $\Phi_e$ threading the loop, we can induce a self-Kerr on the target storage mode, see figure 3b) of the main text.

To implement the self-Kerr gate in a circuit-QED architecture, we consider a SNAIL capacitively coupled to the storage mode:

$$\hat{H} = \omega_s \hat{a}^\dagger \hat{a} + \hat{H}_{\text{snail}} + g\hat{n}_a \hat{n}_s, \tag{S9}$$

where $\omega_s$ is the frequency of the storage mode, $g$ is the coupling between the storage and SNAIL mode, and $n_a, n_s$ are the charge operators corresponding to the storage and SNAIL modes respectively. The SNAIL Hamiltonian is described by

$$\hat{H}_{\text{snail}} = 4E_c(\hat{n} - n_g)^2 + \frac{1}{2}E_L(\hat{\varphi} - \varphi_s[\varphi])^2 + U_s(\varphi_s[\varphi]), \tag{S10}$$

where $E_c, E_L$ are the capacitive and linear inductive energies, respectively, and $\varphi_s[\varphi]$ is the SNAIL phase. The potential energy $U_s$ is given by:

$$U_s(\varphi_s) = -\alpha E_J \cos(\varphi_s) - nE_J \cos\left(\frac{\varphi_{ext} - \varphi_s}{n}\right), \tag{S11}$$

where $E_J$ is the Josephson energy, $\alpha$ is the ratio of the smaller junction to the larger junctions, $n$ is the number of junctions in the longer chain, and $\varphi_{ext}$ is the external flux threading the loop. Finally, the SNAIL phase $\varphi_s[\varphi]$ must satisfy the nonlinear current requirement [4]

$$0 = x_J(M\varphi_s - \varphi) + \alpha \sin(\varphi_s) + \sin\left(\frac{\varphi_s - \varphi_{ext}}{n}\right), \tag{S12}$$

where $x_J = E_L/E_J$. Given that the SNAIL potential is periodic, we perform a Fourier series decomposition of the potential and thus write our Hamiltonian as:

$$\hat{H}_{\text{snail}} = 4E_c(\hat{n} - n_g)^2 - \sum_n c_n(\varphi_{\text{ext}})e^{-in\hat{\varphi}}. \tag{S13}$$

We diagonalize this Hamiltonian, then use the lowest 10 eigenstates to subsequently diagonalize Eq. (S9). Finally, identifying the lowest energy eigenstates of the combined Hamiltonian with maximum overlap of the corresponding bare states permits extraction of the Kerr and cubic Kerr terms. Assuming the hybridized SNAIL remains in the ground state, we then have an effective Hamiltonian:

$$\hat{H}(\Phi_e) = \frac{K(\Phi_e)}{2}\hat{a}^{\dagger 2}\hat{a}^2 + \frac{K_3(\Phi_e)}{6}\hat{a}^{\dagger 3}\hat{a}^3, \tag{S14}$$

where $K_3$ is the 'cubic-Kerr' term. This procedure is repeated at a set of external fluxes to extract the nonlinearities in figure 3b) of the main text. For the SNAIL parameters, we choose $E_C/2\pi = 127$MHz, $E_J/2\pi = 90$GHz, $E_L/2\pi = 150$GHz, $\alpha = 0.07$, and $n = 3$. For the chosen parameters, the SNAIL frequency $\omega_s/2\pi$ changes as a function of the external flux from approximately $4.51 - 5.35$GHz. The storage frequency is chosen to be $\omega_s/2\pi = 8.1$GHz with coupling $g/2\pi = 260$MHz. The choice of a large-detuning ensures a relatively lower cubic-Kerr contribution, at the cost of a decreased Kerr gate rate.

### A. Impact of cubic-Kerr on magic state preparation

We briefly consider the impact of the cubic-Kerr term on the preparation of magic states. In Fig. S7 we consider the action of the Kerr gate on an ideal finite-energy GKP state $|Y_\Delta\rangle$ over a range of cubic-Kerr values, choosing ratios $K/K_3 = 10^2$ to $10^4$ and $\Delta = 0.3$. We plot both the rejection probability after 30 consecutive rounds of sBs and the final post-selected fidelity. The net rotation in phase-space caused by the additional cubic-Kerr term has been accounted for by a frame rotation. We find that, beyond values

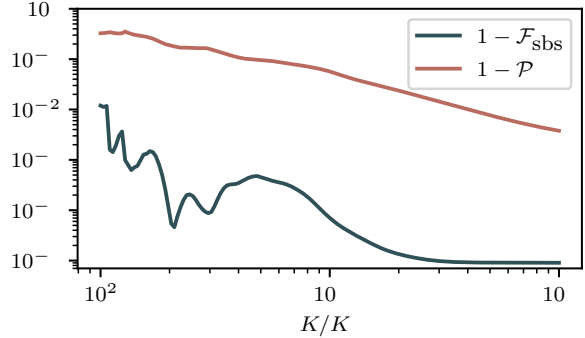

FIG. S7. State infidelity $(1 - \mathcal{F}_{\text{sbs}})$ and rejection probability $(1 - \mathcal{P})$ after applying a self-Kerr gate acting on an ideal finite-energy GKP $|Y_\Delta\rangle$ state for $\Delta = 0.3$, over a range of cubic-Kerr values.

of $K/K_3 > 10^3$, the cubic-Kerr term has very limited impact on the state fidelity, and we are primarily limited by the overlap of the finite-energy GKP states. The SNAIL parameters listed above result in a ratio $K/K_3 = 900$ at the operating point. This ratio can be increased by further optimization of the parameters, increasing the SNAIL/storage detuning, and decreasing the coupling. However, these changes can also decrease the maximum induced value of the self-Kerr $K$, leading to a longer gate which is more susceptible to photon loss. Finding optimal values would naturally become a more complex optimization problem, which we leave to future work.

---

[S1] J. Hastrup, M. V. Larsen, J. S. Neergaard-Nielsen, N. C. Menicucci, and U. L. Andersen, Phys. Rev. A **103**, 032409 (2021).

[S2] F. Hopfmueller, M. Tremblay, P. St-Jean, B. Royer, and M.-A. Lemonde, Quantum **8**, 1539 (2024).
[S3] V. V. Sivak, A. Eickbusch, B. Royer, S. Singh, I. Tsioutsios, S. Ganjam, A. Miano, B. L. Brock, A. Z. Ding, L. Frunzio, S. M. Girvin, R. J. Schoelkopf, and M. H. Devoret, Nature **616**, 50–55 (2023).
[S4] N. E. Frattini, V. V. Sivak, A. Lingenfelter, S. Shankar, and M. H. Devoret, Phys. Rev. Appl. **10**, 054020 (2018).



\end{document}